\documentclass[aps,showpacs,pre,10pt,twocolumn,superscriptaddress]{revtex4-1}

\usepackage{amsmath}
\usepackage{amssymb}
\usepackage{amscd}
\usepackage{wasysym}
\usepackage[ansinew]{inputenc}
\usepackage[T1]{fontenc}
\usepackage{ae,aecompl}
\usepackage{hyperref}
\usepackage{sidecap}
\usepackage{multirow}

\usepackage{color}
\definecolor{red}{rgb}{1,0,0}
\definecolor{blue}{rgb}{0,0,0.8}
\definecolor{green}{rgb}{0,0.5,0}
\definecolor{lred}{rgb}{1,0.7,0.7}
\definecolor{lgreen}{rgb}{0.7,1,0.7}
\definecolor{llred}{rgb}{0.98,0.98,0.98}
\definecolor{llgreen}{rgb}{1,1,1}
\definecolor{shadecolor}{rgb}{1,0.8,0.3}
\definecolor{brown}{rgb}{0.85,0.33,0.10}
\usepackage{colortbl}
   
\usepackage[pdftex]{graphicx}
\DeclareGraphicsExtensions{.pdf}

\newcommand{\be}{\begin{equation}}
\newcommand{\ee}{\end{equation}}
\newcommand{\bea}{\begin{eqnarray}}
\newcommand{\eea}{\end{eqnarray}}
\newcommand{\nn}{\nonumber}

\newcommand{\Hz}{\rm Hz}

\begin{document}
\bibliographystyle{apsrev}

\title{Critical links and nonlocal rerouting in complex supply networks}

\author{Dirk Witthaut}
\affiliation{Network Dynamics, Max Planck Institute for Dynamics and Self-Organization (MPIDS), 
 37077 G\"ottingen, Germany}
\affiliation{Forschungszentrum J\"ulich, Institute for Energy and Climate Research -
	Systems Analysis and Technology Evaluation (IEK-STE),  52428 J\"ulich, Germany}
\affiliation{Institute for Theoretical Physics, University of Cologne, 
		50937 K\"oln, Germany}

\author{Martin Rohden}
\affiliation{Network Dynamics, Max Planck Institute for Dynamics and Self-Organization (MPIDS), 
 37077 G\"ottingen, Germany}
\affiliation{Third Institute of Physics, Faculty of Physics, Georg August University, 37077 G\"ottingen,
  Germany}

\author{Xiaozhu Zhang}
\affiliation{Network Dynamics, Max Planck Institute for Dynamics and Self-Organization (MPIDS), 
 37077 G\"ottingen, Germany}

\author{Sarah Hallerberg}
\affiliation{Network Dynamics, Max Planck Institute for Dynamics and Self-Organization (MPIDS), 
 37077 G\"ottingen, Germany}

\author{Marc Timme}
\affiliation{Network Dynamics, Max Planck Institute for Dynamics and Self-Organization (MPIDS), 37077 G\"ottingen, Germany}
\affiliation{Institute for Nonlinear Dynamics, Faculty of Physics, Georg August University G\"ottingen, 37077 G\"ottingen,  Germany}

\date{\today }

\begin{abstract}
Link failures repeatedly induce large-scale outages in power grids and other 
supply networks. Yet, it is still not well understood, which links are particularly 
prone to inducing such outages. Here we analyze how the nature and 
location of each link impact the network's capability to maintain stable supply. 
We propose two criteria to identify critical links on the basis of the topology 
and the load distribution of the network \emph{prior to} link failure. They 
are determined via a link's redundant capacity and a renormalized linear 
response theory we derive. These criteria outperform critical link prediction 
based on local measures such as loads. The results not only further our 
understanding of the physics of supply networks in general. As both criteria 
are available before any outage from the state of normal operation, they 
may also help real-time monitoring of grid operation, employing 
counter-measures and support network planning and design.
\end{abstract}



\maketitle

% --- content ------------------------------------------------

\section{Introduction}

The robust operation of physical distribution and supply networks is 
fundamental for economy, industry and our daily life.
For instance, a reliable supply of electric power 
fundamentally underlies the function of most of our technical
infra\-structure \cite{Stro01,Amin05,Pour06,Marr08,Krog08,Hine09}. 
In periods of high loads, the breakdown of single infrastructures such as 
transmission lines can cause a global cascade of failures implying large-scale 
outages with potentially catastrophic consequences \cite{Pour06,Albe00,Cohe01,Albe04,Simo08,Buld10,Hine10,12braess,13nonlocal,Helb13}. 
Periods of extreme loads are expected to become more likely in future grids as
power from renewable sources, such as wind turbines, is often generated far
away from the consumers (e.g. off-shore) and moreover strongly fluctuating
\cite{Amin05,Heid10,Mila13,Pesc14}. It is thus crucial to understand which
factors limit robustness of supply networks and in particular to identify
those links that are indispensable for network operation, compare also
\cite{Krog08,Bald08,Bomp09,Schn11,Grad12,Menc14,Scha14}. 
Are there simple characteristics in the physics of complex supply networks 
to identify which of their links are critical? 

Many approaches to identify critical transmission lines in power
grids or critical links in other supply networks fundamentally rely, for instance, on
large-scale numerical simulations of detailed models that, after a local
breakdown, emulate the future dynamics of large parts of the network
\cite{Bald08,Salm04,UCTE04,Kati10}.
Here, we propose a complementary approach to predict \emph{a priori} which
links are critical in the sense that their failure yields larger-scale
malfunction (outage) in the network.
We identify two concepts that reveal, prior to any outage, how the network 
topology jointly with the load distribution influence which links are critical: 
The first relies on the link's redundant capacity which we quantify, the second
originates from a renormalized linear response theory we derive.
Based on these 
theoretical insights, we propose two network-based criteria to identify critical links.
A statistical evaluation suggests that the new
measures predict critical links much more reliably than standard loads or
flows. 

\begin{figure*}[tb]
\centering
\includegraphics[width=12cm]{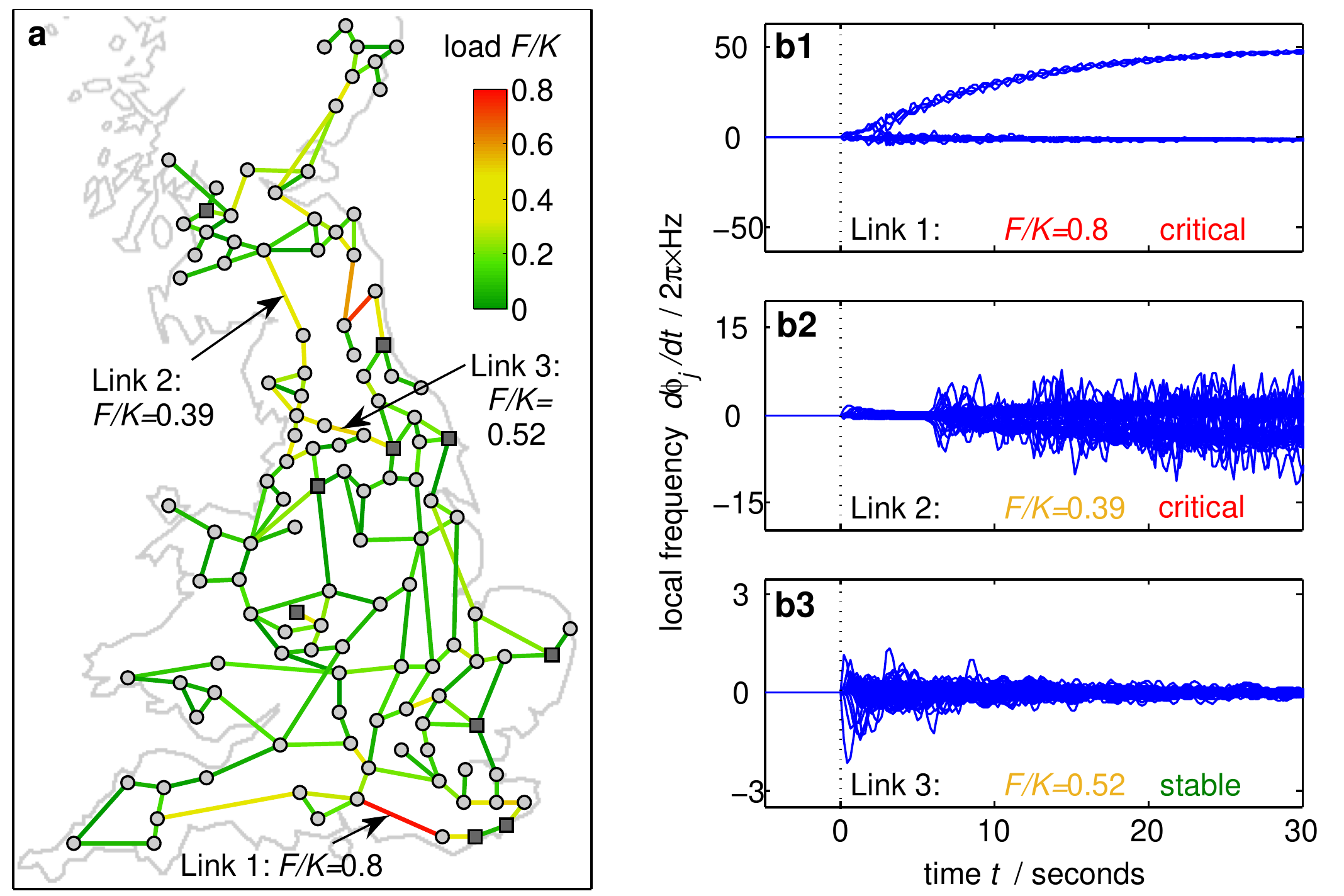}
\caption{
\label{fig:gbexample}
Limits to predicting critical links by local measures. 
(a) Stationary loads in a coarse-grained model of the British 
power grid. For three links  $(a,b)$ 
the loads $L_{ab}=F_{ab}/K_{ab}$ are indicated.
(b) The load 
does not necessarily indicate whether a local failure causes a 
desynchronization and thus a major malfunction.
(b1) Highly loaded link (labeled 'link 1' in panel a) induces 
   desynchronization and is thus critical.
(b2) Moderately loaded link '2' still induces desynchronization 
    and thus is also critical.
(b3) In the same network, even a more heavily loaded link '3' 
   does not yield desynchronization and leaves the network fully functional. 
   Accordingly, link '3' is highly loaded but non-critical (stable).
We analyze critical links for test networks based 
on the topology of the British high-voltage transmission grid 
\cite{Simo08,12powergrid}: Ten nodes are randomly chosen as generators 
($\square, \, P_j = + 11 \, P_0$) and the others as consumers ($\circ, \, P_j = - P_0$), 
with $P_0 = 1\, {\rm s}^{-2}$. Lines connecting to generator nodes have 
a larger transmission capacity $K_{ij} = 2 \, K_0$ 
than the remaining links with $K_{ij} = K_0 = 15 \, {\rm s}^{-2}$. 
See appendices \ref{sec:networkdata} and \ref{sec:additional-res}) 
for additional test networks.}
\end{figure*}

\section{Loads, Flows, and Critical Links}

To obtain insights into mechanisms underlying large-scale outages and to pin
down how the structure of a network determines its vulnerability,  we base our
analysis on a dynamic model of AC power grids
\cite{Berg81,Fila08,12powergrid,Menc14}. Simpler generic models of supply
networks and more complex load flow models of engineering yield qualitatively
the same results, see appendices \ref{sec:gridmodels} and
\ref{sec:additional-res}. The AC model 
characterizes the dynamics of the power angle $\phi_j(t)$ by 
\begin{align} 
       \frac{d^2 \phi_j}{dt^2} = P_j - \alpha \frac{d \phi_j}{dt}
          + \sum_i^N K_{ij}\sin(\phi_i-\phi_j)
        \label{eqn:eom}  
\end{align}
for $N$ synchronous machines $j\in \{1,\ldots,N\}$. Here,
$\phi_j(t)=\theta_j(t)-\Omega t$ is the phase $\theta_j(t)$ of unit $j$
relative to the grid reference phase oscillating with the grid's cycle 
frequency $\Omega$, e.g. $\Omega = 2 \pi \times 50 \, \Hz$, 
$\omega_j = d\phi_j/dt$ 
is the deviation from the grid reference frequency, $P_j$ is the effective 
demand or supply (of a consumer or \emph{sink} where $P_j<0$ or a 
producer or \emph{source} where $P_j>0$, respectively) of unit $j$, 
$\alpha$ is a damping constant and the link capacities $K_{ji} = K_{ij} \geq 0$ 
are proportional to the susceptance of the transmission line $(i,j)$. 

The power flow 
\be
  F_{ij} = K_{ij} \sin(\phi_i - \phi_j),
  \label{eq:powerflow}
\ee 
from unit $i$ to $j$ is then determined via the units' phase difference such
that the (relative) load is given by  $L_{ij}=F_{ij}/K_{ij}=\sin(\phi_i -
\phi_j)$. Stationary dynamics, as relevant on time scales of seconds in normal
operation, is then given by a synchronized ('phase-locked') solution with
temporally fixed phase differences  $\phi_i - \phi_j$
\cite{Stro01,Piko03,Ashw06,Timm07,Zill07,Fila08,Scho08,13powerlong, Dorf13, Mott13,14bifurcation}.

How can we identify which links are critical? The load distribution of a given
supply network in normal stationary operation may serve as a first hint. Due
to the distributed nature of sources and sinks and the topology of the
network, some links are much more loaded than others (Fig.~\ref{fig:gbexample}). 
Intuitively, rerouting the load of a highly loaded link should be harder than 
that of a less loaded one \cite{Ahuj13,Alde13}. It has been
shown that attacks on highly loaded links \emph{on average} have more 
severe consequences than random failures (see, e.g. \cite{Albe04,Hine10}).

Interestingly, whether or not the failure of a particular link induces
network desynchronization is often not predictable by local measures such as load. 
In fact, single link failures may have completely distinct consequences for global 
network operation, largely independent of load:  For instance, whereas the failure of 
one highly loaded link (link 1 in  Fig.~\ref{fig:gbexample} a) may cause a 
desynchonization of phases and thus a
large-scale network outage (see Fig.~\ref{fig:gbexample} b1), the failure of a
 less and only moderately loaded link (link 2) may still induce large-scale outage
 (Fig.~\ref{fig:gbexample} b2). This notwithstanding, the failure of a
 third link  (link 3) that is more highly loaded than link 2, is uncritical to
 network operation (Fig.~\ref{fig:gbexample} b3).
 So predicting outages based on a link's load alone may have substantial limitations.

In what follows, we classify all links of a network into 'critical' ones, those whose failure
induces long-term desynchronization and thus a non-functional network state,
and 'stable' ones, those whose failure leaves the network functional.
In particular, we integrate the the equations of motion (\ref{eqn:eom}) numerically
starting from a stationary state of normal operation (phase locking).
As it turns out, which links are critical depends jointly on the global topological 
structure of a network and its collective dynamics, in particular on the 
link's location within the grid topology and the entire grid's load distribution.

\section{Quantifying Network Redundancy}

For a supply network to remain stable, it needs sufficient options for rerouting 
the (directed) flow $F_{ab}$ originally assigned to a failing link $(a,b)$. 
Therefore, a sufficient degree of redundancy of the remaining network also 
matters to reliably detect critical links. How much redundancy is sufficient? 

\begin{figure*}[t]
\centering
\includegraphics[trim=0cm 0cm 7.6cm 0cm, clip=true, width=12cm, angle=0]{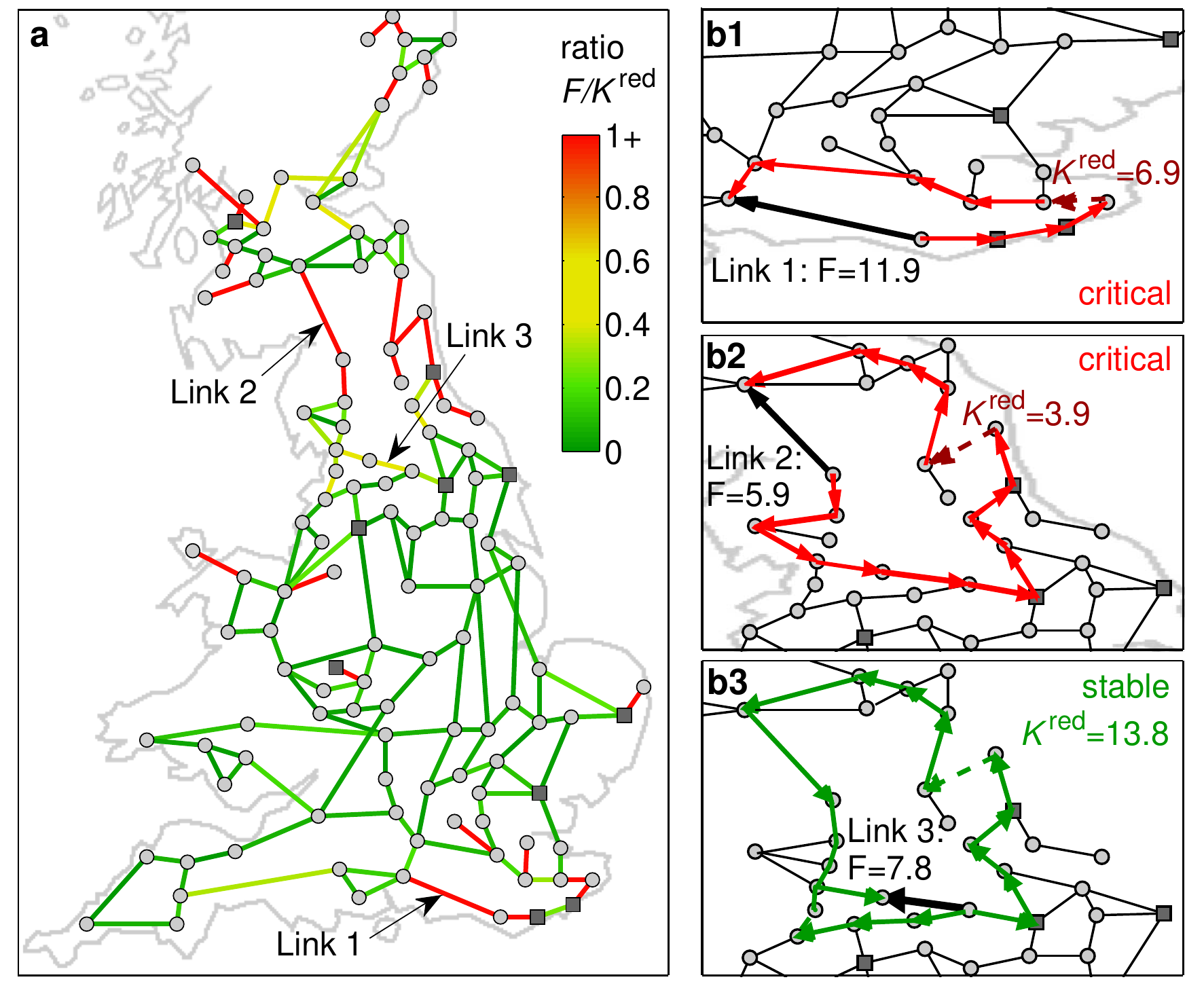}
\caption{\label{fig:RedundancyNotLoad}
Redundant capacity indicates flow rerouting options.
(a) Ratio between flows and redundant capacities (color coded), with
the three links from Fig. \ref{fig:gbexample} indicated by arrows.
(b) Dominant rerouting paths. (b1,b2) For the two links (1,
heavily loaded) and (2, moderately loaded), the flows to be rerouted are
larger than the redundant capacity. 
The bottlenecks on the rerouting path are indicated by dashed red arrows.
Thus $|F_{ab}/K^{\textsf{red}}_{ab}|>1$ and these links are critical. 
(b3)
For link 3, whereas more heavily loaded than link 2, the flow to be rerouted
is smaller than the total available redundant capacity. 
Thus $|F_{ab}/K^{\textsf{red}}_{ab}|<1$ and the link is stable.
}
\end{figure*}

To identify critical links, we quantify the redundancy as follows:
If the link $(a,b)$ fails, 
the (directed) flow $F_{ab}$ has to reroute over alternative paths in the network. 
However, the links along these paths have only a limited residual capacity $K_{ij} - F_{ij}$ 
to take over this flow.  We define the redundant capacity  $K_{ab}^{\rm red}$ of a
link $(a,b)$ as the maximum flow that can be transmitted from unit $a$ to $b$ 
over the residual network excluding link $(a,b)$ (see appendix \ref{sec:kred} 
for an algorithm to determine it).
The redundant capacity is approximately given by
\be
    K_{ab}^{\rm red} = 
     \sum_{\stackrel{\rm paths}{a \rightarrow b}}\min_{(i,j) \in a \rightarrow b}
        (K_{ij} - F_{ij}) \ ,
\label{eqn:Kred}
\ee
that is, 
the minimum residual capacity across the links $(i,j)$ on a path $a \rightarrow b$, 
summed over different alternative paths $a \rightarrow b$  
in the residual network which do not share a `bottleneck', i.e. a link
$(i,j)$ where the minimum in (\ref{eqn:Kred}) is attained. If there is only 
one such path then the expression becomes exact.
We thus propose to consider the ratio of the actual flow and the redundant
capacity as a measure to predict critical links as
\bea
   &&  |F_{ab}/K_{ab}^{\rm red} |  > h  \;  \Rightarrow \;  
                       \mbox{predicted to be critical}, \nn \\
   &&  |F_{ab}/ K_{ab}^{\rm red}|  \le h  \; \Rightarrow \;  
                    \mbox{predicted to be stable}, 
   \label{eqn:RedundantCapacityPredictor}
\eea
where $h$ is a threshold value that can be optimized for any given
specific prediction.
Figure \ref{fig:RedundancyNotLoad} compared to Fig.\ref{fig:gbexample}
illustrate that predictions based on redundant capacity (\ref{eqn:RedundantCapacityPredictor}) 
may work even for links where load-based predictions fail.

\section{Flow Rerouting in Linear Response}
Alternatively, we analyze how general alterations of the capacity of a single link modify the 
global operation of a network. Consider a small perturbation $\kappa_{ij}$ of the 
network capacities at a single link $(a,b)$ such that $K'_{ij}  = K_{ij} + \kappa_{ij}$ 
with $\kappa_{ab} = \kappa_{ba} = \kappa$ and $\kappa_{ij} = 0$ for all other 
links. This perturbation induces a change  $\phi_j \rightarrow \phi'_j$ 
of the steady state phases of the network. 

\begin{figure}[tb]
\centering
\includegraphics[width=8cm, angle=0]{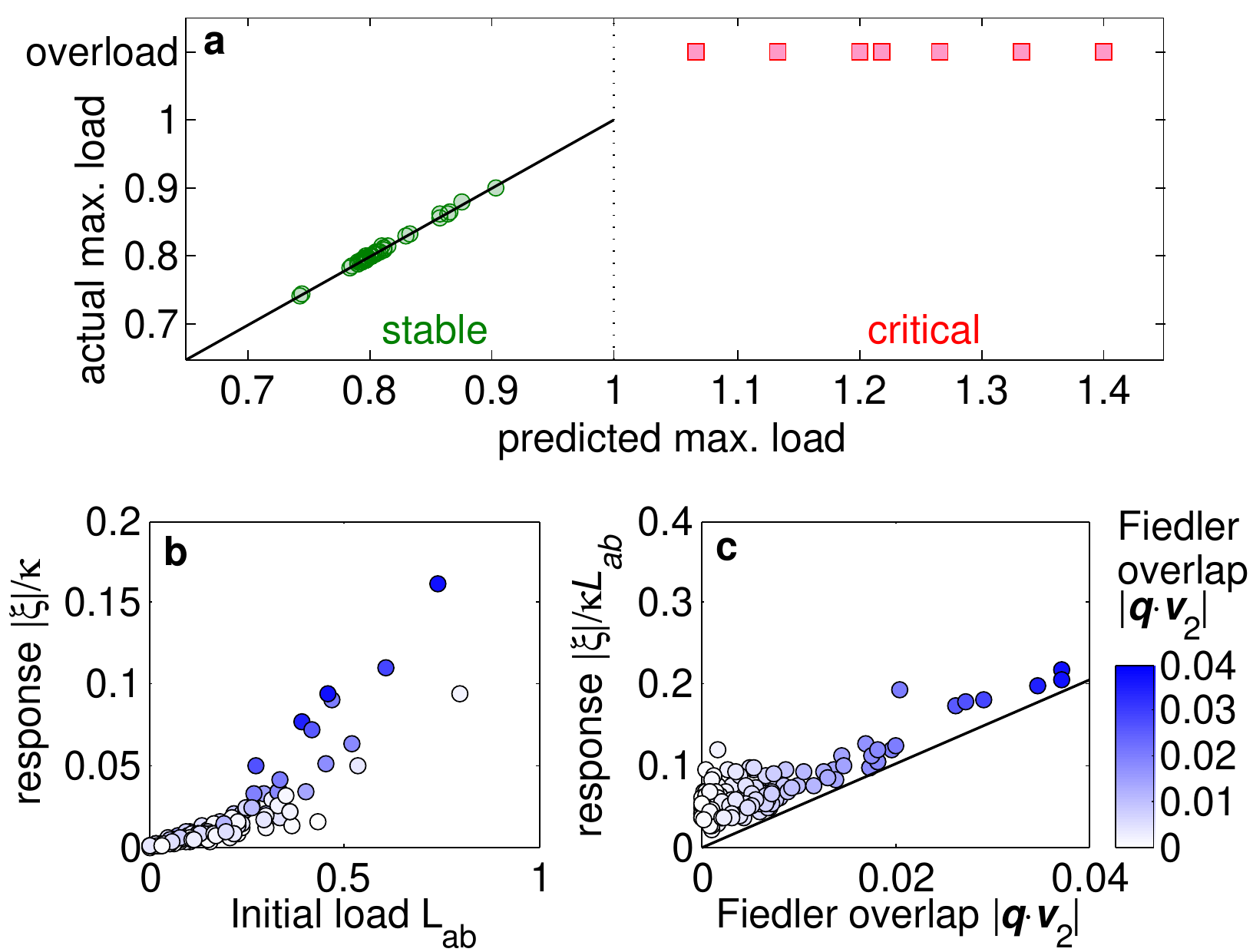}
\caption{\label{fig:LinearResponsePredictor}
Renormalized linear response theory predicts critical links.
(a) Renormalized linear response not only predicts which links are
critical (red squares) and which are stable (green disks), it also accurately
tells the maximum loads for stable links (solid line, no fit parameter). 
(b) Response as a function of the load $L_{ab}$ of
the failing link does grow proportionally to the load, but also depends on
other factors. 
(c) Response grows proportionally to the overlap
$|\boldsymbol{q} \cdot  \boldsymbol{v}_2|$ with
the Fiedler vector. 
The solid straight line with slope slope $1/\lambda_2$ gives a strict lower bound
see Eq.~(\ref{eqn:LinearResponseSecondEV}). Networks and parameters as in 
Fig.~\ref{fig:RedundancyNotLoad}.
% The response of a given network is characterized by the 
%norm of the phase difference $\boldsymbol{\xi}$ and the columns
%of the susceptibility $\eta_{\cdot\leftarrow ab }$.
%In addition, the phase difference is significantly enhanced if the 
%overlap to the Fiedler vector 
%$|\boldsymbol{q} \cdot  \boldsymbol{v}_2|$ is large.
%(c,d) The global response, averaged over all links 
%is proportional to the inverse algebraic connectivity $1/a_2$.
%Shown is the slope of the phase reponse and the matrix norm 
%of the susceptibility $\eta$ for networks with different values 
%$K_0=10,12,\ldots,40 \, {\rm s}^{-2}$ and thus different connectivity.
%Network topology as in figure \ref{fig:predictionLinearResponse}.
% plotting: aclinks_sus3.m and aclinks_sus4.m 
}
\end{figure}

Expanding the perturbed steady state ($\ddot \phi'_j = \dot \phi_j' = 0$ in 
Eq.~(\ref{eqn:eom})) to first order in the response $\xi_j := \phi_j' - \phi_j$ 
and subtracting the unperturbed steady state condition yields 
\be    
   \sum_i K_{ij} \cos(\phi_i-\phi_j)(\xi_i-\xi_j) = - \sum_i\kappa_{ij}\sin(\phi_i-\phi_j).
   \label{eq:differencePerturbedOriginal}
\ee
During normal operation, we can assume that $|\phi_i - \phi_j|\leq \pi/2$
for all links in the network \cite{14bifurcation} and thus use (\ref{eq:powerflow})
to obtain the identity 
$
   K_{ij} \cos(\phi_i -\phi_j) 
   = \sqrt{K_{ij}^2 - F_{ij}^2} =:  \widetilde{K}_{ij} \, .
$
This quantity characterizes the available capacity that the network may use to 
respond to perturbations in terms of the network topology and the original flows.
We define the Laplacian matrix \cite{Newm10} 
$
   \Lambda_{ij} := -  \widetilde{K}_{ij} + \delta_{i,j} \sum \nolimits_n
   \widetilde{K}_{i n},
   \label{eqn:Laplacian}
$
of this available capacity using the Kronecker-delta symbol ($\delta_{x,y}=1$ if $x=y$ 
and $\delta_{x,y}=0$ otherwise).
With the vectorial components $q_i= (\delta_{i,a}-\delta_{i,b})$, equation 
(\ref{eq:differencePerturbedOriginal}) is recast into the form
$      
   \Lambda \boldsymbol{\xi} = \kappa L_{ab}\; \boldsymbol{q} \, .
   \label{eqn:matrix}
$
We remark that this equation is linear but depends nonlinearly on the system's
unperturbed state variables $\phi_i$. Using the Moore-Penrose pseudo-inverse $T :=
\Lambda^{+}$, yields the phase responses
\be
   \boldsymbol{\xi}  = \kappa L_{ab} T  \boldsymbol{q} \ . 
     \label{eqn:matrix-inv}
\ee
Thus the power flow (\ref{eq:powerflow}) in the perturbed network 
is to first order in $\kappa$ given by 
\bea
   F'_{ij} &=& (K_{ij} + \kappa_{ij}) \sin( \phi_i - \phi_j + \xi_i - \xi_j ) \nn \\
    &=& F_{ij} +  \kappa \, \eta_{ij \leftarrow ab} \, ,
\label{eq:FlowReroutingSmall}
\eea
where we have defined the \emph{link susceptibility}
$
   \eta_{ij \leftarrow ab} :=
     L_{ab} \times [ \widetilde K_{ij} (T_{jb} - T_{ja} - T_{ib} + T_{ia}) 
        + (\delta_{ia} \delta_{jb} - \delta_{ja} \delta_{ib}) ].
$
So if capacities are perturbed slightly, 
eqn.~(\ref{eq:FlowReroutingSmall}) tells us, to first order in the 
perturbations $\kappa$, which flows increase, which decrease and 
how much.

\begin{figure}[tb]
\centering
%\includegraphics [height=6cm, angle=0]{fig4abc_gbhet400_helv14.pdf}
%\hspace{1cm}
\includegraphics[width=8cm, angle=0]{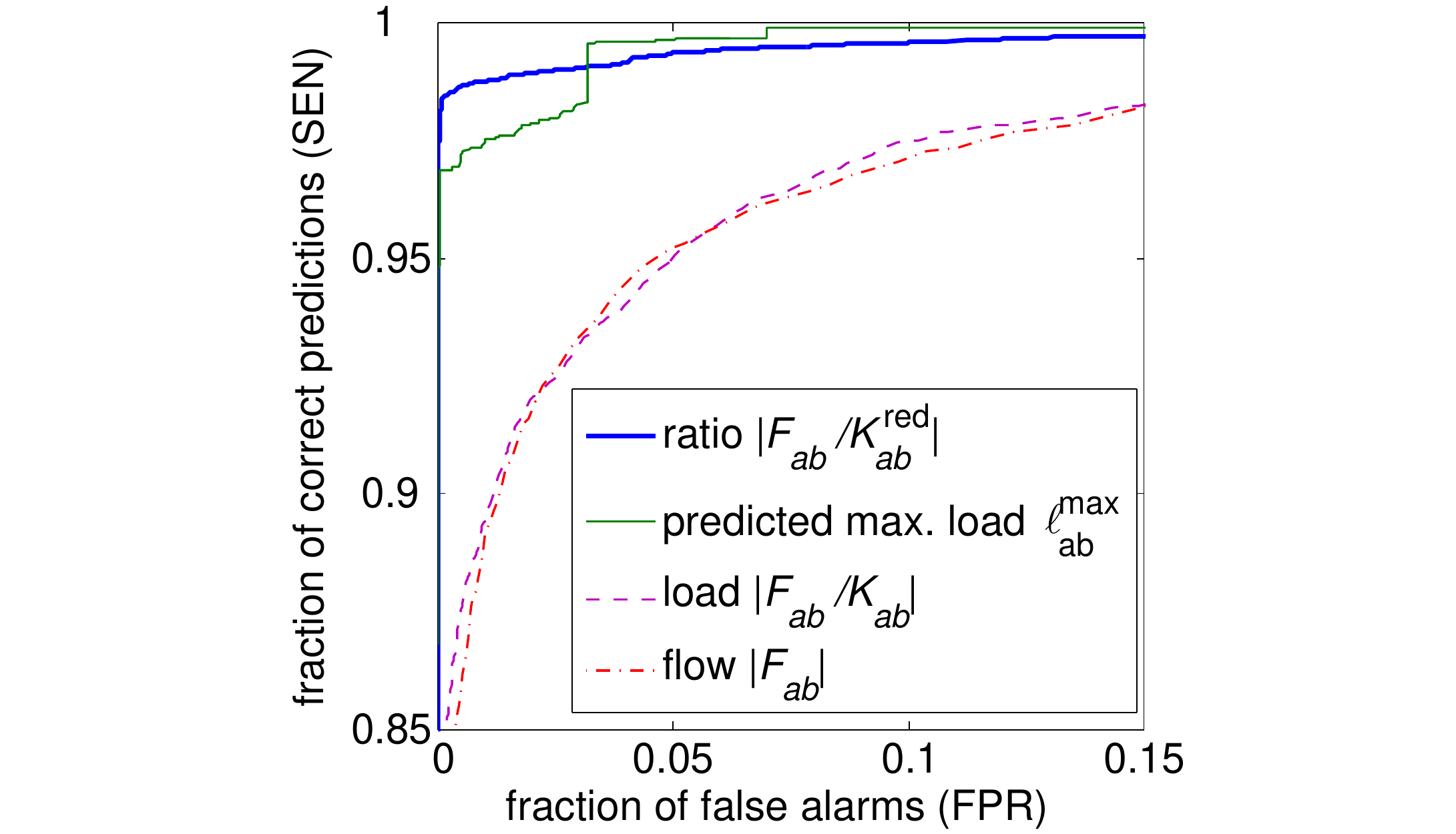}
\caption{\label{fig:hist1}
Improved quality of critical link prediction.
%(a-c)
%Probability distributions conditioned on the predictors
%(a) loads, $|F_{ab}/K_{ab}|$,
%(b) flows relative to the redundant capacities, $|F_{ab}/K_{ab}^{\textsf {red}}|$, and
%(c) maximum loads from renormalized linear response theory,
%$\ell ^{\max }_{ab}=\max _{(i,j)} |F''_{ij}/K_{ij}|$ for critical (red) and stable
%(green) links, respectively. The stronger differences between the
%distributions for stable and critical edges already indicate the redundant
%capacity predictors and linear response predictors are more reliable than
%predictors based on loads only.
% (d) 
A Receiver operating characteristic curves indicate a multi-fold increase in prediction 
quality using the classifiers (\ref{eqn:RedundantCapacityPredictor}) and 
(\ref{eqn:LinearResponsePredictor}) in comparison to load or flow.
(Statistics for 66000 links from 400 random grid realizations; model, topologies 
and parameters as in Figure \ref{fig:gbexample})
}
\end{figure}

Even for perturbations $\kappa$ that are not small, this linear response argument 
reliably predicts \emph{where} the flow is rerouted (see appendix \ref{sec:linres} for 
a detailed analysis).
To accurately also predict the \emph{magnitude} of the flow change for a large 
perturbation of total link failure, we self-consistently renormalize the linear response 
by replacing 
$
\kappa := -F_{ab}/\eta_{ab \leftarrow ab} \ ,
$
thereby ensuring that the flow of the defective link vanishes, $F'_{ab} = 0$. 
From this renormalized linear response theory, the modified estimates for the
rerouted network flows read
\be
   F''_{ij} = F_{ij} - \frac{\eta_{ij \leftarrow ab}}{\eta_{ab \leftarrow ab}}
              \, F_{ab}.
  \label{eqn:flow-linres}
\ee
The $F''_{i,j}$ do of course depend on the considered link $(a,b)$, that for 
clarity does not explicitly appear as an index.
Related measures are used in engineering where they are 
referred to as line outage distribution factors \cite{Wood13}.

As the rerouted total flows on each link must not be larger than the 
respective link capacities, we propose the \emph{predicted maximum load}
$\max_{(i,j)} |F''_{ij}/K_{ij}|$ as a discriminating measure between critical 
and stable links:
\bea
   && \max_{(i,j)} |F''_{ij}/K_{ij}| >    h   \;  \Rightarrow \;  \mbox{predicted to be critical}, \nn \\
   && \max_{(i,j)} |F''_{ij}/K_{ij}| \le h   \; \Rightarrow \;  \mbox{predicted to be stable}.
   \label{eqn:LinearResponsePredictor}
\eea
In addition, every bridge is predicted to be critical.  

This renormalized linear response theory does not only well identify which
links are critical and which are stable, based on criterion
(\ref{eqn:LinearResponsePredictor}); the fundamental equation
(\ref{eqn:flow-linres}) also predicts the value of the expected
maximum load across the network (Fig.~\ref{fig:LinearResponsePredictor}a).

\section{Load and Topology Co-Act}
  
The response $\boldsymbol{\xi}$ as explicated by (\ref{eqn:matrix-inv}) and 
thus the susceptibility $\eta$ are directly proportional to the load $L_{ab}$ of a link.
The load of a link thus does determine its relevance, but other collective
characteristics of the network are equally important.

For heavy loads, some of the responsive capacities $\widetilde{K}_{ij}$ tend
to zero and the associated network becomes weakly connected. The
second eigenvalue $\lambda_2$ of the Laplacian $\Lambda$, which
measures the algebraic conenctivity becomes small (while $\lambda_1\equiv0$, 
independent of the network and operating state) \cite{Newm10}.
Then the pseudo-inverse $T=\Lambda^+$ is dominated by $1/\lambda_2$
and the response is given by 
\be
   \boldsymbol{\xi} \approx \frac{\kappa L_{ab}}{\lambda_2} 
       (\boldsymbol v_2 \cdot \boldsymbol q)  \boldsymbol v_2 \ ,
   \label{eqn:LinearResponseSecondEV}
\ee
up to terms of order $1/\lambda_3$. Here, $\boldsymbol v_2$ denotes 
the eigenvector associated with $\lambda_2$.
Thus the response is specifically also determined by two intrinsically
collective properties of the load distribution: the inverse algebraic
connectivity  $1/\lambda_2$ and the overlap 
$(\boldsymbol v_2 \cdot \boldsymbol q)$, which is large for 
links connecting different components of the network 
\cite{Fort10,si}.
Direct numerical simulations confirm these theoretical 
predictions, Fig.~\ref{fig:LinearResponsePredictor} b,c.

\section{Quality of Critical Link Predictions}
The two proposed classifiers (\ref{eqn:RedundantCapacityPredictor}) and
(\ref{eqn:LinearResponsePredictor}) enable the prediction of critical links
with high accuracy. To test the performance we generate 400 random
network realizations varying the generator positions, thereby analyzing 66000
links in total. A quantitative assessment of the different
classifiers is provided by a receiver operating characteristic
(ROC) curve, Fig.~\ref{fig:hist1}, displaying the
fraction of correct predictions of critical links (the sensitivity) as a function of the
fraction of false alarms (the false positive rate) when the discrimination threshold $h$ is varied.
We see that the two classifiers introduced in this paper closely approach the perfect 
operating point $(0,1)$ with almost no false alarms and alsmost perfect sensitivity
(Fig.~\ref{fig:hist1}).
The total number of incorrect predictions, including false alarms and missed
critical links, is as small as 395 (0.6 percent) for the the linear response measure 
and 920 (1.4 percent) for the linear response measure.
Thus it is possible to reduce the incorrect predictions based on loads  
(3149, 4.7 percent) and flows (2937, 4.5 percent) by a factor of more 
than seven, thereby drastically improving performance. 
Additional macroscopic quantifiers such as those based on the area under 
the ROC-curve confirm this view for a variety of different network topologies
(see appendix \ref{sec:additional-res})

\section{Discussion}
In summary, we link the overall network topology with the load distribution
resulting from the collective network dynamics and present
non-local relations to identify a network's response to
link failures.  On this basis, we propose two network-based strategies to identify critical
links by (i) quantifying the redundant capacity of the network and by (ii)
estimating the flow rerouting through developing a renormalized linear response
theory. The analysis suggests that the two proposed predictors are well suited
to identify critical links. In particular they provide a substantial improvement
of the quality of predicting critical links compared to predictions based on local 
measures such as a link's load. We emphasize that the quality of prediction is 
evaluated by statistical measures whereas the predicting measures are derived from 
insights about the topological connectivity and the nonlinear dynamics of the networks.

By construction, the proposed criteria readily generalize to other network
topologies, to the physics of generic linear supply networks and to complex
load flow models as standard in engineering (see appendices \ref{sec:gridmodels} and
\ref{sec:additional-res}).
For network operation, the proposed predictors may provide key hints for initial
analysis, because they suggest which links require special attention, e.g. in
detailed large scale simulations of power grid engineering. 
In particular, our insights about collective influences indicate that the weak 
spots of a network are not necessarily given by the most heavily loaded elements 
(cf.~\cite{Albe04,Hine10}) and that network-based rather than local measures 
provide suitable guidelines for the security assessment of real-world power 
grids (cf.~\cite{Bald08}).

\begin{acknowledgments}

We thank Moritz Matthiae for valuable discussions and Peter Menck 
for providing network data. We gratefully acknowledge support from 
the Federal Ministry of Education and Research (BMBF grant no.~03SF0472E 
to M.T. and D.W.), the Helmholtz Association (grant no.~VH-NG-1025 to D.W.) 
and the  Max Planck Society to M.T.

\end{acknowledgments}

\appendix

\section{Alternative Models and Methods}

\subsection{Models for power grid operation}
\label{sec:gridmodels}

In this section we review different models for the simulation and analysis 
of power grids.  

\subsubsection{Power flow in AC grids}

In an AC power grid the voltages and currents oscillate sinusoidally with period 
$T$ and angular frequency $\omega = 2\pi/T$,
\bea
   U(t) &=& U_\text{peak} \cos(\omega t + \theta_U) = \sqrt{2}\, \Re( \underline U e^{i \omega t}  ), \nn \\
    I(t) &=& I_\text{peak} \cos(\omega t + \theta_I) =\sqrt{2} \, \Re( \underline I e^{i \omega t}  ),
\eea
where we have defined the \emph{complex}-valued amplitudes
\bea
    \underline U = \frac{U_\text{peak}}{\sqrt{2}}  e^{i \theta_U}, \qquad
     \underline I = \frac{I_\text{peak}}{\sqrt{2}}  e^{i \theta_I}.
\eea

The electric power is oscillating, too. The net transferred energy per period is called 
the real or active power, 
\bea
   P &=& \frac{1}{T} \int_0^T U(t) I(t) =  \Re(\underline U \, \underline I^*).
\eea
The imaginary part $Q = \Im(\underline U \, \underline I^*)$ describes the power which is 
temporarily stored in capacities or inductances and is referred to as the reactive power. 
Furthermore, one defines the apparent power
$S = P+ i Q = \underline U \, \underline I^*$. These relations show
that the relative phases of voltages and currents are essential for the power 
flow in a grid. Calculations are performed most easily using the complex 
notation defined here. In the following we drop the underline to keep the notation
simple.

Consider a power grid with $k,\ell = 1,\ldots,N$ nodes. The basic relation between 
voltage at the nodes and the currents between them is Ohm's law,
\be
   I_{k \ell} = \frac{1}{Z_{k \ell}} (U_k - U_\ell) = 
         y_{k \ell} (U_k - U_\ell),
\ee
where $Z_{k \ell}$ is the impedance and $y_{k \ell} = 1/Z_{k \ell}$ the admittance of the 
transmission line between the nodes $k$ and $\ell$.. 
The admittance is divided into its 
real part, the conductance, and its imaginary part, the susceptance,
\begin{equation}
   y_{k \ell}= g_{k \ell} + i b_{k \ell}. 
   \label{eq:admittancetwo}
\end{equation}
In case there are multi-circuit transmission lines we assume that $y_{k \ell}$ is the sum of
the admittances of the single circuits and $y_{k \ell}$ equals zero if no transmission
line exists. The total current injected to the node $k$ is then given by
\be
   I_k = \sum_\ell I_{k \ell} = \sum_\ell y_{k \ell} (U_k - U_\ell).
\ee
For actual calculations it is convenient to introduce the \emph{nodal admittance matrix}
 $\mathbf{Y}$,
\begin{equation}
  Y_{k \ell} = G_{k\ell} + iB_{k\ell}= \left\{ 
   \begin{array}{lll}
   \displaystyle\sum \nolimits_{n=1}^{N} y_{k n} &  \mbox{if} & k = \ell; \\ 
     - y_{k \ell} & \mbox{if} & k \neq \ell,
   \end{array} \right.
   \label{eq:admittance}
\end{equation}
Ohm's law can then be rewritten in a vectorial form, which yields the \textit{network equation}
\begin{equation}
   \vec{I} = \mathbf{Y} \;\vec{U} \label{eq:networkeq}
\end{equation}
with $\vec{I} = ( I_1, I_2, ..., I_N )$ representing the \emph{complex} currents injected into 
$N$ nodes, $\vec{U} = ( U_1, U_2, ..., U_N)$ denoting the \emph{complex} voltages.

The apparent power injected to a node $k$ is then given by
\bea
   \label{eq:nodeS}
   S_k &=& U_k I_k^{\ast}  \\
   &=& \sum_{\ell=1}^{N} |U_k||U_\ell| \left(G_{k \ell}-iB_{k \ell}\right) 
      \left(\cos\theta_{k \ell}+i\sin\theta_{k \ell}\right), \nn     
\eea
where $\theta_{k \ell}=\theta_k-\theta_\ell$ denotes the difference in phase of the 
complex voltages $U_k$ and $U_\ell$. The effective and reactive power are then
again given by the real and imaginary part of this expression \cite{Grai94,Wood13},
\begin{eqnarray}
   P_k &=&\displaystyle\sum_{\ell=1}^{N} |U_k||U_\ell|
            \left(G_{k \ell}\cos\theta_{k \ell}+B_{k \ell}\sin\theta_{k \ell}\right),   
   \vspace{0.3cm}\label{eq:activepower}\\
   Q_k &=&\displaystyle\sum_{\ell=1}^{N} |U_k||U_\ell|
            \left(G_{k \ell}\sin\theta_{k \ell}-B_{k \ell}\cos\theta_{k \ell}\right).
   \label{eq:reactivepower}
\end{eqnarray}

\subsubsection{An oscillator model for power grid operation}

We model the power grid as a network of $N$ rotating machines
representing, for instance, wind turbines, or electric motors. Each machine 
$k = 1,\ldots,N$ is characterized by the mechanical power $P_k^{\rm mech}$ 
acting on it, which is positive for a generator and negative for a consumer. 
The state of a machine is determined by its angular frequency and the rotor 
or power angle $\theta_k(t)$ relative to the reference axis rotating at the
nominal grid angular frequency  $\omega_0 = 2\pi \times 50$ Hz or 
$\omega_0 = 2 \pi \times 60$ Hz, respectively. 
Correspondingly, $\omega_k = d \theta_k / dt$ gives the angular frequency
deviation from the reference $\omega_0$. The dynamic of the rotor is governed 
by the \emph{swing equation} \cite{Prab94,Mach08,Berg81,Fila08,12powergrid}
\be
  \label{eq:swing}
   M_k \frac{d^2 \theta_k}{dt^2} + D_k \frac{d \theta_k}{dt}
    = P_k^{\rm mech} - P_k^{\rm el},
\ee
where $P_k^{\rm el}$ is the electrical power that is transmitted to or from 
other rotating machines.
The parameter $M_k$ is the moment of inertia of the rotor times $\omega_0$ 
and $D_k$ measures the damping, which is mainly provided by damper
windings. If a generator is decoupled from the grid in an uncontrolled way 
such that $P_k^{\rm el} = 0$, it accelerates until finally an emergency 
shut-down is performed.

For simplicity we here neglect ohmic losses in the grid such that
the line admittance is purely imaginary, $Y_{k \ell} = i B_{k \ell}$. Furthermore,
we assume that the magnitude of the voltage is constant throughout
the grid, $|U_k| = U_0$ for all nodes $k=1,\ldots,N$. For a common two-pole
synchronous machine the phase of the voltage equals the mechanical 
phase of the rotor. 
The expression  (\ref{eq:activepower}) for the active electric power 
then simplifies to
\be
   P_k^{\rm el}  =  \sum_{\ell=1}^N  U_0^2 B_{k \ell} \sin (\theta_{k} - \theta_\ell). 
\ee
Substituting this result into the swing equation yields
the equations of motion
\be
   M_k \frac{d^2 \theta_k}{dt^2} + D_k \frac{d \theta_k}{dt}
    = P_k^{\rm mech}  - \sum_{\ell=1}^N  U_0^2 B_{k \ell}  \sin (\theta_{k} - \theta_\ell). 
   \label{eqn:eom-theta}
\ee
Using the abbreviations 
\bea
     P_k &=& \frac{P_k^{\rm mech} - D_k \omega_0}{M_k}, \qquad
     \alpha_k = \frac{D_k}{M_k},  \nn \\
    K_{k\ell} &=& \frac{U_0^2 B_{k\ell}}{M_k}, \nn
\eea
the oscillator model reads
\begin{equation}
   \frac{d^2 \theta_k}{dt^2} = P_k - \alpha_k \frac{d \theta_k}{dt}
          + \sum_{\ell=1}^N K_{k\ell} \sin(\theta_\ell - \theta_k) \, .
        \label{eqn:osc-model}  
\end{equation}
These equations of motion are equivalent to the structure-preserving model 
\cite{Berg81} in power engineering. The model is mathematically equivalent, but 
derived from slightly different assumptions. Notably, the model is also very similar 
to the so-called Kuramoto model, which is studied extensively in statistical 
physics \cite{Kura75,Stro00,Aceb05}. In the Kuramoto model the inertia term is 
absent such that it can be viewed as the over-damped limit case.

\subsubsection{Including voltage dynamics}
\label{sec:3rd-order}

The oscillator model can be extended to include the dynamics of the local
voltages, which yields the so-called third-order model \cite{Prab94,Mach08,Schm13}.
The equations of motion for the mechanical motion of the 
synchronous machines is still given by equation (\ref{eqn:osc-model})
but the coupling strength becomes time-dependent via the local voltages
\be
    K_{k\ell}(t) = \frac{B_{k \ell }}{M_k} \, U_k^{(q)}(t) U_\ell^{(q)}(t).
      \label{eq:3rd-capacity}
\ee

The equations of motion for the voltages can be derived from Faraday's and
Ohm's law applied to the voltages, currents and fluxes of the stator and
rotor windings of a synchronous generator. Assuming again lossless transmission
lines one obtains
\bea
   T_k \frac{d U_k^{(q)}}{dt} 
       &=& U_k^{(f)}  - U_k^{(q)} - \Delta X_k^{(d)} I_k^{(d)} \nn \\
       &=& U_k^{(f)}  - U_k^{(q)} - \Delta X_k^{(d)} \sum_{\ell = 1}^N B_{k \ell} \cos \theta_{k \ell},
   \label{eq:volt-dyn}
\eea
where $U_k^{(f)}$ is the voltage of the field windings and $\Delta X_k^{(d)}$ is
determined by the transient and static reactances of the stator windings. The actual 
derivation of this formula can be found in the literature
\cite{Prab94,Mach08,Schm13} such that we do not go into detail here. 
Instead we briefly discuss the mathematical structure of the equation.

\begin{figure}[tb]
\centering
\includegraphics[clip=true, trim= 8.5cm 9cm 5cm 6cm, width=6cm, angle=0]{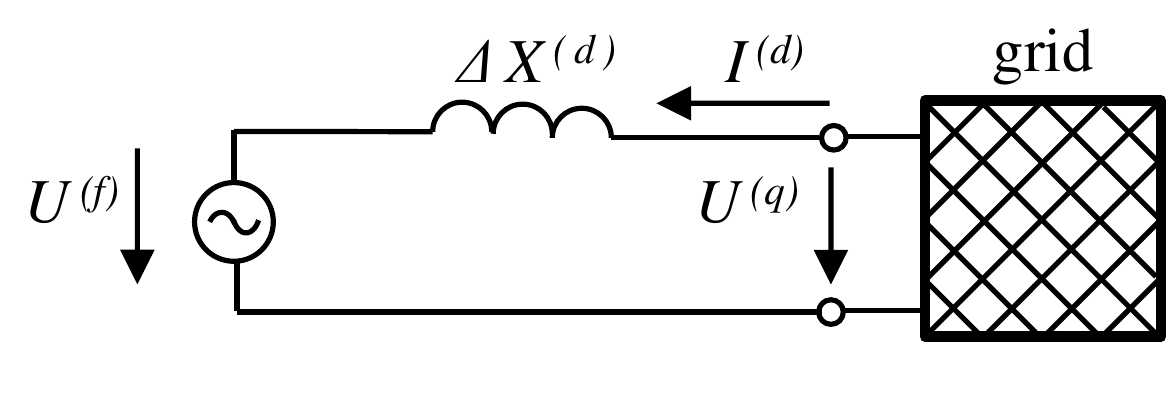}
\caption{\label{fig:3rd-equivalent}
{Equivalent circuit diagram explaining the internal voltage 
drop in a synchronous machine.}
}
\end{figure}

If the phase differences $\theta_{k \ell}$ would be constant in time, then the 
equation of motion (\ref{eq:volt-dyn}) would just describe an exponential 
relaxation to the equilibrium value
\be
   U_k^{(q)} = U_k^{(f)}  - \Delta X_k^{(d)} \sum_{\ell = 1}^N B_{k \ell} \cos \theta_{k \ell},
\ee  
which can be understood in terms of the equivalent circuit diagram shown in 
Fig.~\ref{fig:3rd-equivalent}.  
The equilibrium voltage $U_k^{(q)}$ seen by the grid is given by the source 
voltage $U_k^{(f)}$ reduced by the internal voltage drop over the reactance 
$\Delta X_k^{(d)}$.    
The relaxation time $T_k$ is given by the ratio of the inductance and the 
ohmic resistance of the field windings as in a resonant circuit.
     
However, the phase differences $\theta_{k \ell}$ are not constant in time but 
again depends on the voltages. This gives rise to a complex interplay of phase 
and voltage dynamics and new potential instabilities as discussed in
\cite{Schm13}.

\subsubsection{Load flow calculations}
\label{sec:load-flow}

Load flow calculations are routinely carried out in power engineering. They include 
ohmic losses and reactive power but are generally static. Hence they describe the 
steady state operation of a power grid but not the dynamics of a major outage. 

In load flow calculations a system of algebraic equations describing the complex power 
flows through the transmission lines in the network is established. Such an equation 
system can be formulated when the network topology, the amount of power supply 
and consumption at each node, as well as the admittance of each transmission line 
are known. Solving the equation system numerically, one obtains information about 
the steady state operation of a power grid.
Critical states of a power grid occur when either the equation system does not have a solution 
or the load of a transmission line exceeds a limiting value which depends on the line parameters. 
Such an overload generally does not lead to an immediate desynchronization of the grid 
but to an overheating or bending on longer time scales and eventually to an emergency shut 
down.

In the following we present a short derivation of system of equations that represents the 
load flow model \cite{Grai94,Wood13}. Within the load flow model we distinguish three types of nodes: 
generator buses (PV buses), load buses (PQ buses) and slack buses. PV buses are connected 
to generators, therefore their active power $P$ and their voltage magnitudes $|U|$ are fixed. 
PQ buses are connected to loads, i.e. their net injected active power $P$ and reactive power 
$Q$ are given, while the voltages at these nodes are unknown. A special kind of bus is the slack 
bus. It does not exist in reality, but is introduced to facilitate numerical computation. 
Active power $P$ and reactive power $Q$ of the slack bus remain unspecified in order to 
balance the active and reactive power in the system during the iteration towards the steady 
state solution. The physical reason for this is that the power losses on the transmission lines 
are not known \textit{a priori}, i.e. before the solution is obtained. The slack bus acts as an
ideal voltage source and ensures the power balance of the grid. Table \ref{tab:bustype} 
summarizes given and unknown variables that characterize the different types of buses.

\begin{table}[th]
\begin{center}
		\renewcommand{\arraystretch}{1.4}
		\setlength{\tabcolsep}{0pt}
		  \begin{tabular}{!{\vrule width 2pt}c!{\vrule width 1pt}c!{\vrule width 1pt}c!{\vrule width 1pt}c!{\vrule width 1pt}c!{\vrule width 2pt}}
			\noalign{\hrule height 2pt}
											&	\large{$P$}		& \large{$Q$}  	& \large{$|U|$}		& \large{$\theta$}\\  \noalign{\hrule height 1pt}
			PV bus& known &\cellcolor{shadecolor}\,unknown\,& known				& \cellcolor{shadecolor}\,unknown\, \\ \noalign{\hrule height 1pt} 
			PQ bus& known & known				& \cellcolor{shadecolor}\,unknown\,&\cellcolor{shadecolor}\,unknown\, \\ \noalign{\hrule height 1pt} 
			\,Slack bus\, &\cellcolor{shadecolor}\,unknown\,& \cellcolor{shadecolor}\,unknown\, & known				& known\\ \noalign{\hrule height 2pt}
		  \end{tabular}
\caption{
Characteristics of the three types of buses in load flow calculation.
}
\label{tab:bustype}
\end{center}
\end{table}

Suppose we are interested in a power transmission network of $m$ generators and $n$ loads.
The unknown state variables are the phases of all $n+m$ nodes of the network and the voltages
of the $n$ loads summarized in the state vector
\be
   \vec x = \left(\theta_1,\cdots,\theta_{n+m},U_1,\cdots,U_{n}\right).
\ee  
Using all the known information (see Table \ref{tab:bustype}), we obtain a system of 
$2 n  + m$ equations describing the active power $P$ and the reactive power $Q$ 
injected to the buses (see Eq.~\ref{eq:activepower} and Eq.~\ref{eq:reactivepower})
\begin{eqnarray}
\label{eq:pf}
      P_1\left(\theta_1,\cdots,\theta_{n+m},U_1,\cdots,U_{n}\right)&=&P_1^{\,\text{sp}} \nonumber \\
      \vdots && \vdots \nonumber\\
      P_{n+m}\left(\theta_1,\cdots,\theta_{n+m},U_1,\cdots,U_{n}\right)&=&P_{n+m}^{\,\text{sp}} \nonumber\\
\\
      Q_1\left(\theta_1,\cdots,\theta_{n+m},U_1,\cdots,U_{n}\right)&=&Q_1^{\,\text{sp}} \nonumber\\
      \vdots&&\vdots \nonumber\\
      Q_{n}\left(\theta_1,\cdots,\theta_{n+m},U_1,\cdots,U_{n}\right)&=&Q_{n}^{\,\text{sp}}. \nonumber
\end{eqnarray}
The superscript \textquotedblleft sp\textquotedblright\ to the variables on the right-hand side 
indicates that the values of this variables are specified beforehand.
We thus have a system of $2 n  + m$ nonlinear algebraic equations for 
$2 n  + m$ unknown variables $\vec x$.

A common and effective way to solve this equation system is to use the Newton-Raphson 
method \cite{Prab94}, which iteratively updates the state vector $\vec x$ from an initial 
guess towards the solution of Eqs.~(\ref{eq:pf}). The final solution for the state vector 
describes the state of the power transmission network in normal, steady-state operation.

\subsubsection{The DC approximation and linear flow models}

When the loads and losses in a power grid are small, then the load flow calculations can 
be simplified considerably \cite{Grai94,Wood13,Purc05,Hert06}. Within the so-called 
DC approximation one first neglects all ohmic losses such that the admittance of a 
transmission line is purely imaginary, $G_{k\ell} = 0$. The real power injected at 
node $k$ then reads (cf.~equations (\ref{eq:activepower}) )
\be
   P_k =\sum_{\ell = 1}^N B_{k\ell} |U_k| \, |U_\ell|
                       \sin(\theta_k - \theta_\ell). \nn
         \label{eq:dc-power}         
\ee
In a second step one assumes that the magnitude of the voltages in the grid are 
fixed, typically at the reference value of the respective voltage level. Technically, all 
nodes must then be considered as PV buses, such that no equations for the reactive 
power must be taken into account when solving the load flow equations. Finally, 
small loads imply that the phase differences across a transmission line are small 
such that the sine function can be approximated to first order as
\be   
     \sin( \theta_k - \theta_\ell ) \approx  
          \theta_k - \theta_\ell .
\ee
The load flow calculations then reduce to a set of linear equations
\be
   \sum_{\ell = 1}^N \widetilde K_{k\ell} \theta_\ell = P_k^\text{sp},
   \label{eq:dcapprox}
\ee  
where the coupling coefficients are defined as 
\begin{equation}
  \widetilde K_{k\ell}= \left\{ 
   \begin{array}{lll}
    \displaystyle \sum \nolimits_{n=1}^{N} B_{k n}  |U_k| |U_n| & 
             \mbox{if} & k = \ell; \\ [2mm]  
     - B_{k \ell} |U_k| |U_\ell| & \mbox{if} & k \neq \ell.
   \end{array} \right.
\end{equation}

The simplified equation (\ref{eq:dcapprox}) is referred to as the DC approximation, 
as it is mathematically equivalent to Kirchhoff's circuit equation for a DC electric circuit. 
Still, it describes the flow of real power in AC power grids. Obviously, linear equations 
can be solved much faster than the nonlinear load flow equations (\ref{eq:pf}). The 
DC approximation is particularly advantageous when the flow must be calculated for 
many different scenarios of generation and load, because the matrix $\widetilde K$ 
has to be inverted only once. The limits of its applicability are discussed in detail 
in \cite{Purc05,Hert06}.

Notably, a mathematically equivalent model describes the flow in hydraulic 
networks \cite{Dura04} or vascular networks of plants \cite{Kati10}. In this 
case the model is derived as follows. We denote the flow from node $k$ to 
node $\ell$ by $F_{k \ell}$. The continuity equation then reads
\be
   \sum_{\ell=1}^{N} F_{k \ell}  = P_k \qquad 
   \text{for all } k \in \{1, \ldots , N \},
   \label{eqn:energycon2}
\ee
where $P_k$ is the source or sink strength at node $k$. The unique steady state is then 
determined by the condition that the total dissipated power
\be
    E_{\rm diss} =  \sum \nolimits^{'}_{k<\ell} \frac{F_{k \ell}^2}{2 K_{k \ell}}
\ee
should be minimal.
In this expression, $K_{k \ell}$ denotes the capacity of the link $(k,\ell)$, which is 
determined by the cross section in hydraulic of vascular networks. The primed sum 
runs only over existing transmission lines, i.e. only over links with $K_{k\ell} \neq 0$. 
Minimizing this expression using the method of Lagrangian multipliers yields 
\be
   F_{k\ell} = K_{k\ell} (\theta_k - \theta_\ell),
\ee
where the values of the $\theta_\ell$ are determined by a linear set of equations as in equation 
(\ref{eq:dcapprox}) with the Laplacian matrix
\begin{equation}
  \widetilde K_{k\ell}= \left\{ 
   \begin{array}{lll}
    \displaystyle \sum \nolimits_{n=1}^{N} K_{k n}  & 
             \mbox{if} & k = \ell; \\ [2mm]  
     - K_{k \ell}  & \mbox{if} & k \neq \ell.
   \end{array} \right.
\end{equation}

\subsection{Network Topologies}

\label{sec:networkdata}

\begin{figure}[b]
\centering
\includegraphics[width=8cm]{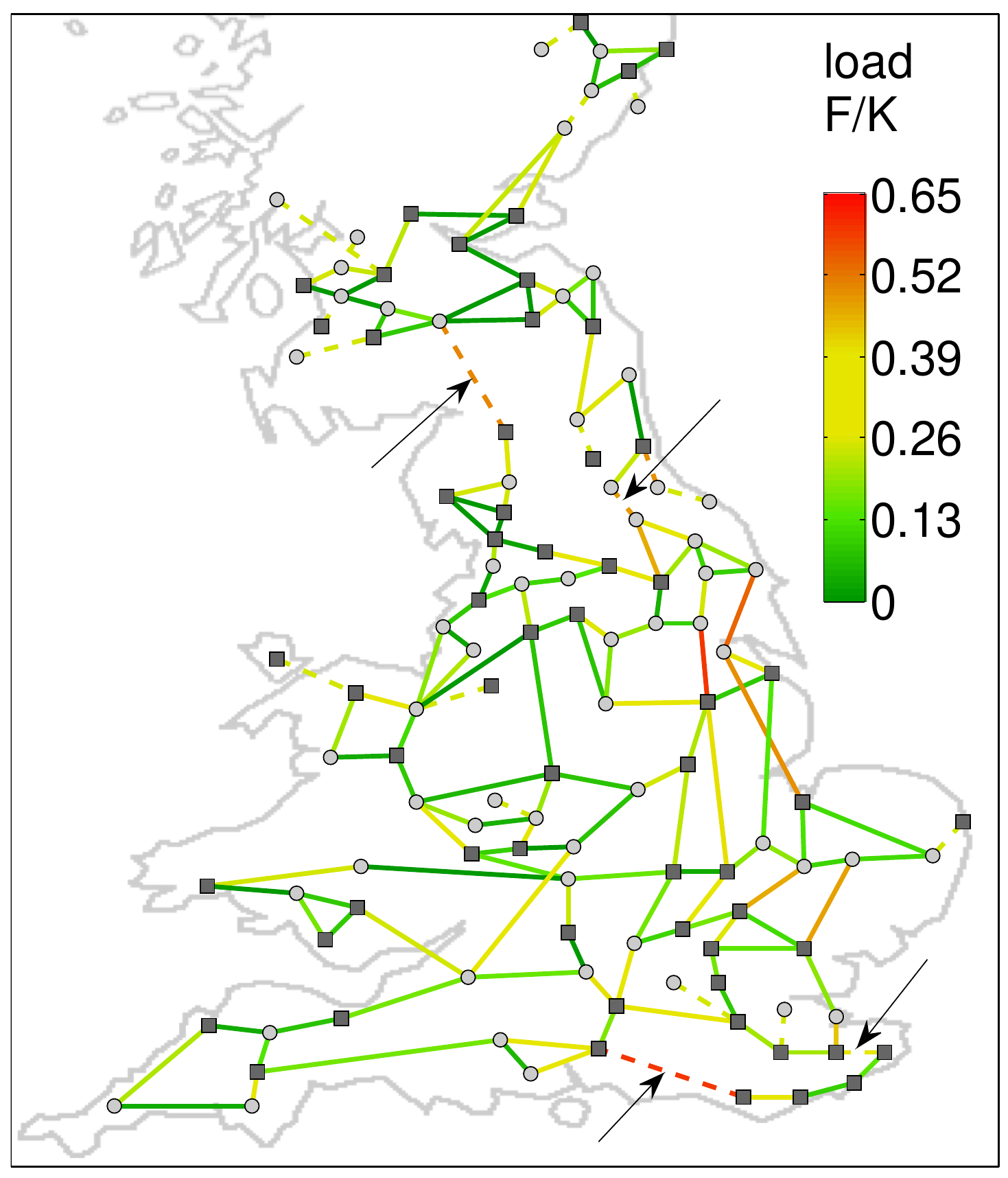}
\caption{\label{gbhommap} 
{Model power grid based on the topology of the 
British high-voltage power transmission grid \cite{Simo08,12powergrid}.} 
Half of the nodes are randomly chosen as generators ($\square$)
and consumers ($\circ$), respectively, with $P_j = \pm 1 \, {\rm s}^{-2}$ (homogenous case).
In the current example 13 bridges and 4 further links are critical (marked by arrows). All of
them are faithfully identified using the novel predictors proposed in this manuscript.}
\end{figure}

\begin{figure}[t]
\centering
\includegraphics[width=8cm]{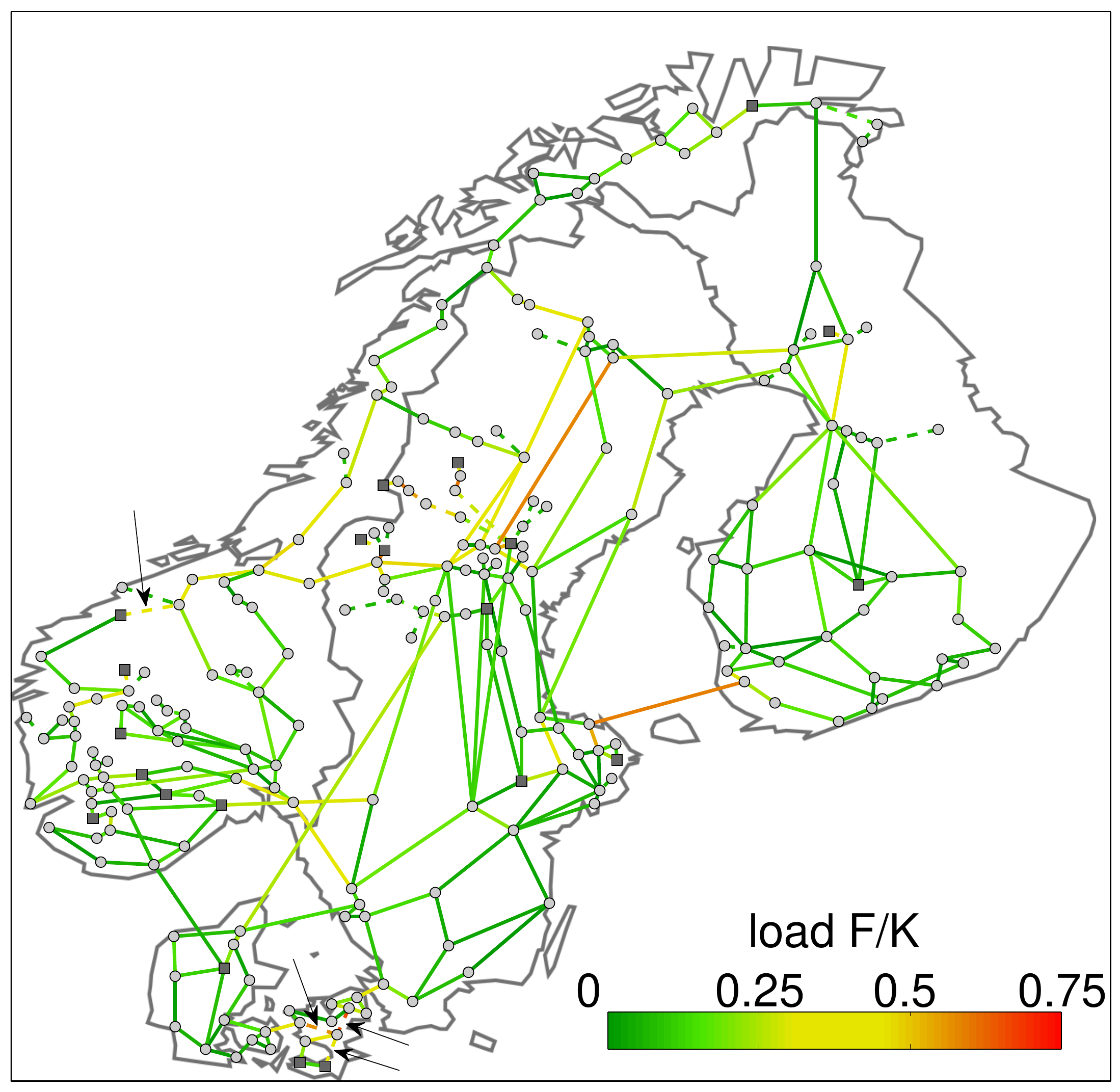}
\caption{\label{scandmap}
{Model power grid based on the topology of the Scandinavian
high-voltage power transmission grid  \cite{Menc14}.} 21 nodes are randomly chosen as 
generators ($\square$), the remaining ones are consumers ($\circ$) (heterogeneous case).
In the current example 45 bridges and 4 further links are critical (marked by arrows). All of
them are faithfully identified using the novel predictors proposed in this manuscript.
}
\end{figure}

We analyze the effects of transmission line modifications and breakdowns for partly synthetic 
model networks based on the topology of different real-world power grids.
We consider the topology of the current high-voltage power transmission grid in
Great Britain with 120 nodes and 165 links \cite{Simo08,12powergrid} (cf.~Fig.~\ref{gbhommap})
and Scandinavia with 236 nodes and 320 links \cite{Menc14} (cf.~Fig.~\ref{scandmap}).

\begin{figure}[t]
\centering
\includegraphics[width=8cm]{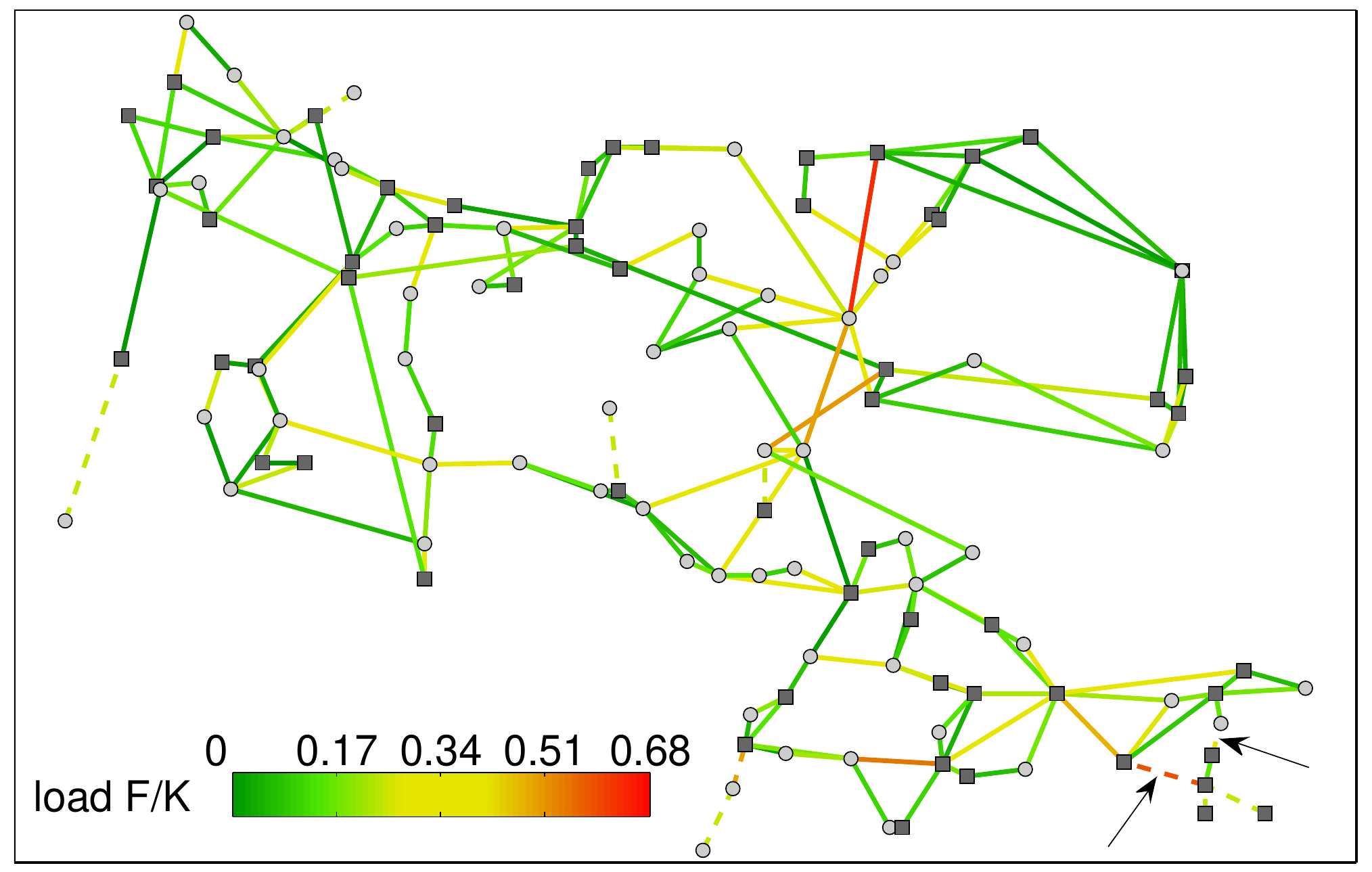}
\caption{\label{ieeemap}
{Model power grid based on the topology of the IEEE
118 bus test grid \cite{powertest}.} Half of the nodes are randomly chosen as generators ($\square$)
and consumers ($\circ$), respectively, with $P_j = \pm 1 \, {\rm s}^{-2}$ (homogeneous case).
In the current example 9 bridges and 2 further links are critical (marked by arrows). All of
them are faithfully identified using the novel predictors proposed in this manuscript.
 } 
\end{figure}

Furthermore we consider the topology of the IEEE 118-bus test grid with 118 nodes and 
179 links (cf.~Fig.~\ref{ieeemap}), representing a part of the electric power system in 
the Midwestern United States as of 1962 \cite{powertest}. We then generate a large 
number of synthetic networks as a basis for statistical analysis by randomizing generation 
and load as described below.

We note that parameters are chosen to represent a \emph{heavily loaded} power grid.
Periods of such high loads are still rare nowadays, but are expected to become much more
likely in future power systems with a large number of strongly fluctuating 
renewable power sources (see, e.g., \cite{Pesc14}). In such situations the grid becomes 
vulnerable to the breakdown of single transmission lines as discussed in the present paper.
A prime example is the European power blackout on 4th November 2006 \cite{UCTE07}.
A large amount of electric power from wind turbines was transmitted from northern 
to south-western Europe resulting in heavy loads of several lines. The planned shut-down
of one double-circuit line over the river Ems in Germany was then sufficient to bring
the entire grid to the tipping point of a breakdown. 
In other words, this grid did not possess enough
redundant capacity $K^{\rm red}$ to securely take over the load of the shut-down line.
A coupling of the busbars at the transformer station Landesbergen then finally
triggered the blackout, in which the European grid fragmented into three asynchronous
areas.

We have performed extensive simulations of the oscillator models for the three
topologies described above for two different scenarios of power generation. 
First, we consider a heterogeneous case, where several large power plants are 
connected to the grid. We randomly select $N_{\rm gen}$ nodes to be the
connection points of the generators, the remaining $N_{\rm con}$ nodes are 
consumers. 
The effective power of the consumer nodes is assumed to be  $P_{\rm con} = - P_0$,
while the generator nodes have $P_{\rm gen} = + N_{\rm con}/N_{\rm gen} \times P_0$.
The links adjacent to the generator nodes are assumed to have a higher
transmission capacity ($2 \times K_0$), while  the remaining links have 
a transmission capacity $K_0$. In all numerical examples we set 
$P_0 = 1 \, {\rm s}^{-1}$ and $K = 15 \, {\rm s}^{-1}$ unless stated otherwise. 
In particular, we set $N_{\rm gen}  = 10$ for the topology of the British grid 
(Fig.~\ref{gbhommap}) and the  IEEE 118 bus test grid (Fig.~\ref{ieeemap}), 
whereas $N_{\rm gen}  = 21$ for the Scandinavian grid (Fig.~\ref{scandmap}).
This choice ensures that $P_{\rm gen}$ is approximately equal in all cases such that
the results are comparable.

Second, we consider a homogeneous case assuming a more 
distributed power generation \cite{12powergrid,Menc14}.
In this case we assume that half of the nods are generators $P_{\rm gen} = + P_0$ and
half are consumers with  $P_{\rm con} = - P_0$. Again the positions of generators and 
consumers are chosen randomly. All links have the save transmission capacity,
which we set to  $K_0 = 4 \, {\rm s}^{-1}$ (British grid and IEEE grid) and
$K_0 = 5 \, {\rm s}^{-1}$ (Scandinavian grid), respectively.
We note that the flow is directly proportional to the load for such a homogeneous
network such that both quantities have the same predictive power.

In section \ref{sec:add-topologies-results} we collect the numerical results for the three model
networks with both heterogeneous and homogeneous generation. 
In all cases the new methods for the predictions of critical infrastructures
introduced in the main manuscript significantly outperform predictors
based on load, flow, topological edge-connectivity or edge betweenness centrality. 
Notably, the linear response
prediction works even better in the case for homogeneous networks
while the redundant capacity approach performs better for heterogeneous 
networks. The difference becomes most apparent for the model grids based on the
topology of the Scandinavian power grid. The numerical results thus fully 
confirm the findings discussed in the main manuscript.

\subsection{Definition and calculation of the redundant capacity}
\label{sec:kred}

The redundant capacity $K_{ab}^{\rm red}$ of a link 
$a \rightarrow b$ is defined as the maximum \emph{additional} 
flow that the network can transmit from node 
$a$ to $b$ when the link $(a,b)$ fails. 
Here, the maximum is defined in a purely graph-theoretical
way, i.e. we search for a flow which satisfies  $|F_{ij}| \le K_{ij}$
for all $i,j$ and flow conservation at every node. 
This definition is purely structural and can thus be applied to every
model of a supply network. It does \emph{not} take into account 
whether such a flow is dynamically possible or stable for the 
particular network type under consideration. 

Furthermore, we note that the redundant capacity
is \emph{directed}, i.e.  $K_{ab}^{\rm red} \neq K_{ba}^{\rm red}$.
This is immediately clear from the fact that the flow is directed.
If an link $(i,j)$ is fully loaded in the sense that $F_{ij} = - F_{ji} = K_{ij}$,
then it cannot transmit any additional flow from node $i$ to node
$j$. However, it can transmit additional flow in the other direction 
from $j$ to $i$. In fact, this would lower the net flow over the
link $(i,j)$ and thereby the load of the link. 

The redundant capacity of the link $(a,b)$ is calculated 
using a modification of the Edmonds-Karp algorithm \cite{Jung12}.
As discussed above, $K_{ab}^{\rm red} \neq K_{ba}^{\rm red}$ 
such that  the algorithm works with directed graphs.\\

\begin{tabular}{p{1.4 cm} p{6.5cm}}

Input: & Capacity matrix $K$, Initial flow matrix $F$, link $a \rightarrow b$\\

Output: & Redundant capacity $K_{ab}^{\rm red}$ of the link 
$a \rightarrow b$. \\

Step 1: & Delete the link $(a,b)$ from the effective network:
$K_{ab}, K_{ba}, F_{ab}, F_{ba} \leftarrow 0$.  \\

Step 2: & Initialize $K_{ab}^{\rm red} \leftarrow 0$. \\ 

Step 3: & Calculate the residual capacity matrix $K^{f} \leftarrow K - F$.\\

Step 4: & Construct a shortest path $p$ from $a$ to $b$ in the directed 
graph defined by the matrix $K^f$. \\

Step 5:  & If there is no path from $a$ to $b$: STOP. \\

Step 6: & Calculate the maximum available capacity along the path $p$:
$K^{f}_{\rm max} \leftarrow   \max_{(i \rightarrow j) \in p}  K^{f}_{ij}$. \\

Step 7: & Add the  maximum available capacity to the redundant capacity:
$K_{ab}^{\rm red} \leftarrow K_{ab}^{\rm red} + K^{f}_{\rm max}$.\\

Step 8: & Increase the flow along the path $p$: 
$F_{ij} \leftarrow F_{ij} +  K^{f}_{\rm max}$ for all links $(i \rightarrow j) \in p$. \\

Step 9: & GOTO step 3. \\

\end{tabular}
\vspace{3mm}

This definition has been formulated for the oscillator model, which is mainly 
studied in the present paper. It can be directly applied to the third-order model
introduced in section \ref{sec:3rd-order}. In this case, the capacity $K_{ij}$ 
is given by equation (\ref{eq:3rd-capacity}) using the equilibrium voltages. 

The concept of the redundant capacity for the prediction of overloads can be
easily adapted to load flow calculations (see section \ref{sec:load-flow}). 
One then has to specify whether power or current flows or both can lead
to dangerous overloads and use these quantities in the algorithm introduced
above. The results are discussed in detail in section \ref{sec:res-loadflow}.

\subsection{Linear response theory}
\label{sec:linres}

\begin{figure*}[tb]
\centering
\includegraphics[height=7cm]{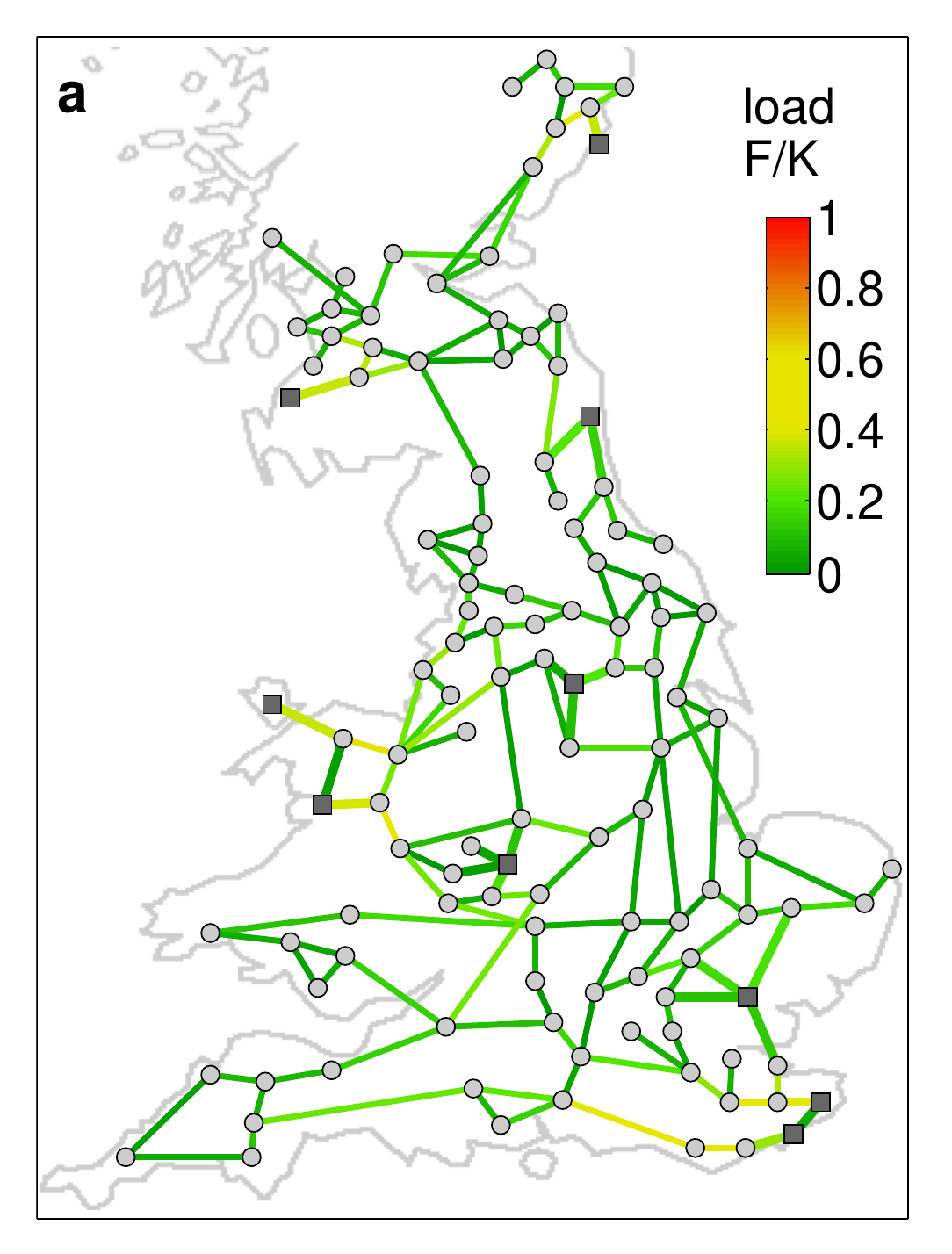}
\hspace{1cm}
\includegraphics[height=7cm]{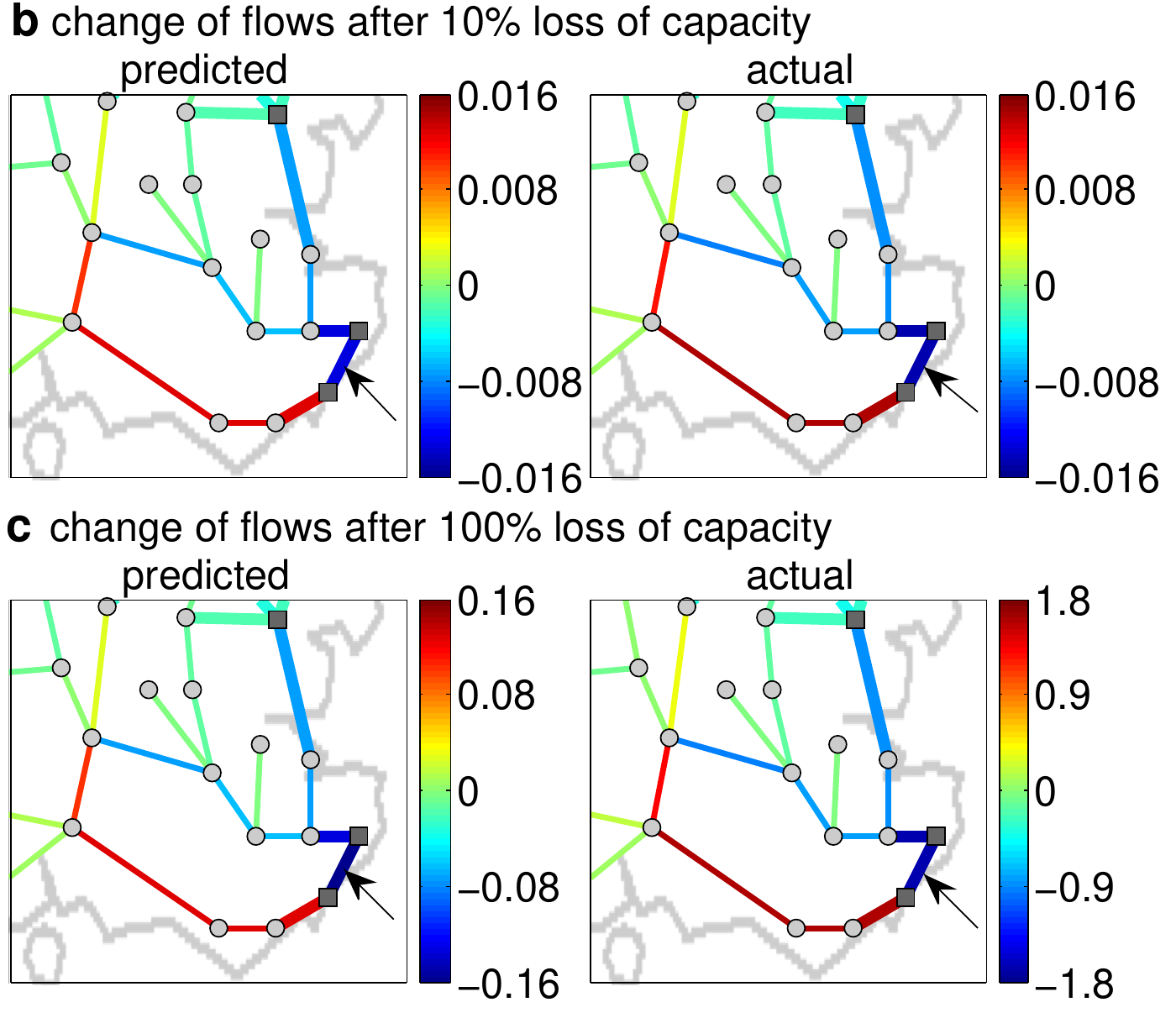}
\caption{
\label{fig:flowchange}
{Prediction of flows by original and renormalized linear response theory.}
(a) Transmission line loads in the steady state in a  
coarse-grained model of the British high-voltage 
transmission grid. 
(b,c) Change of flows $|F_{ij}'| - |F_{ij}|$ in the south-eastern
part of the grid after a perturbation. We analyze (b) the loss of
10 \% of the transmission capacity and (c) the complete breakdown 
of the marked link.
The prediction by linear response theory (\ref{eqn:flow-perturbed})
on the left-hand side is compared to the actual value obtained from
solving the steady state condition for the original and the perturbed
network on the right-hand side. Note the different color scales.
% Simulation and plotting: critical_predict2.m
}
\end{figure*}

Here we discuss the linear response theory for the prediction of
phases and flows in the oscillator model after a local perturbation in more detail.
We consider a small perturbation of the network, 
$K'_{ij} = K_{ij} + \kappa_{ij}$, at a single link $(a,b)$:
\be
  \kappa_{ij} = \left\{
   \begin{array}{l c l}
    \kappa & \;  \mbox{ for } \; &  (i,j) = (a,b) \, \mbox{and} \, (i,j) = (b,a) \\
    0 & & \mbox{all other links.} \\
   \end{array} \right.         
   \label{eqn:allkappa}
\ee
This perturbation induces a small change of the steady state phases of the network 
$\phi_j \rightarrow \phi'_j$. We note that the steady state is defined only up to a 
global phase shift which ha no physical significance. We thus fix the global phase as
$\sum \nolimits_j \phi_j = 0 = \sum \nolimits_j \phi_j'$ and define
the response $\xi_j = \phi'_j - \phi_j$.

The steady state of the original network is defined via
\be
   0 = P_ j + \sum_{i=1}^N K_{ij} \sin(\phi_i - \phi_j)  
   \label{eqn:def-steady}
\ee
for all $j= 1, \ldots, N$.
For the perturbed network we have
\be
   0 = P_ j + \sum_{i=1}^N (K_{ij}+\kappa_{ij}) 
     \sin(\phi_i - \phi_j + \xi_i - \xi_j). 
      \label{eqn:steady-original}
\ee
Expanding the flows to first order in  $\kappa_{ij}$ and the 
response $\xi_j$ yields
\bea
    &&  (K_{ij}+\kappa_{ij})  \sin(\phi_i - \phi_j + \xi_i - \xi_j)   \nn \\
    && \qquad =  K_{ij} \sin(\phi_i - \phi_j) +  \kappa_{ij} \sin(\phi_i - \phi_j) \nn \\
   && \qquad \quad + K_{ij}\cos(\phi_i-\phi_j) (\xi_i-\xi_j) .   
   \label{eqn:steady-perturbed}
\eea
Subtracting the equations  (\ref{eqn:steady-perturbed}) and
(\ref{eqn:steady-original}) then yields
the steady state condition to
leading order in $\xi_j$
with the result
\bea
   0 &=& \sum_{i=1}^N K_{ij}\cos(\phi_i-\phi_j) (\xi_i-\xi_j)+     
      \sum_{i=1}^N \kappa_{ij} \sin(\phi_i-\phi_j)    \nn \\
     &=& \sum_{i=1}^N K_{ij}\cos(\phi_i-\phi_j) (\xi_i-\xi_j) +     
       \kappa L_{ab}  (\delta_{jb} - \delta_{ja})
   \label{eqn:steady2}
\eea
for all $j=1,\ldots,N$.
In the last step we have used the definition of the load
$L_{ab} = F_{ab}/K_{ab} = \sin(\phi_a-\phi_b)$ 
and the perturbation matrix (\ref{eqn:allkappa}).
This result can be further simplified and analyzed by 
 introducing  the matrix
\be
    \Lambda_{ij} := \left\{
    \begin{array}{l c l}  
       - K_{ij} \cos(\phi_i - \phi_j) & \;  \mbox{for} \; & i \neq j \\
      + \sum_\ell  K_{\ell j} \cos(\phi_\ell - \phi_j)    & & i=j. \\
    \end{array} \right.
\ee
and the vector $\vec q$ with the components
\be
   q_j = \left\{
   \begin{array}{l l}
   1  & \; \mbox{for } j = a \\
   -1 & \; \mbox{for } j = b \\
   0  & \; \mbox{else}. \; 
   \end{array} \right.         
\ee
In a short-hand vectorial notation,  equation (\ref{eqn:steady2}) then reads
\be
  \Lambda \vec \xi = \kappa L_{ab}   \vec q.
  \label{eqn:Kxiq}
\ee

Before we go into detail, we first have a closer look at the matrix
$\Lambda$. A steady state of the power grid model (\ref{eqn:eom-theta})
is dynamically stable if the relation $|\phi_i - \phi_j| \le \pi/2$ holds
for all links $(i,j)$ of the network \cite{Mach08,14bifurcation}. Stable 
steady states which do not satisfy this relation exist only at the 
boundary of the stable parameter region. We can thus assume that 
during normal operation we always have  $|\phi_i - \phi_j| \le \pi/2$
for all links such that we can use the following relations:
\bea
    \cos(\phi_i-\phi_j) &=& \sqrt{1 - \sin(\phi_i-\phi_j)^2 }  \ge  0\nn \\
  \Rightarrow K_{ij}\cos(\phi_i-\phi_j) &=& \sqrt{K_{ij}^2 - F_{ij}^2}.
\eea
The expression 
\be
    \widetilde K_{ij} := \sqrt{K_{ij}^2 - F_{ij}^2}
     \label{eqn:kred-def}
\ee
will be referred to as the \emph{responsive capacity} of each link, i.e. 
the free capacity which can be used to react to the perturbation. 
In this case the non-diagonal entries of the matrix $\Lambda$ are 
all non-positive such that $\Lambda$ is a \emph{Laplacian matrix}
for which many properties are known \cite{Newm10}.
In particular, the eigenvalues of a Laplacian matrix satisfy 
$0 = \lambda_1 \le \lambda_2 \le \cdots \le \lambda_N$, where $\lambda_2$ is 
an algebraic measure for the \emph{connectivity} of the
underlying network  \cite{Fied73, Fort10}. We will make use
of these relations to further analyze the properties of the
matrix $\Lambda$ in section \ref{sec:prop-Laplace}.

The Laplacian matrix $\Lambda$ is singular as $\lambda_1 = 0$. Still,
Eq.~(\ref{eqn:Kxiq}) can be solved for $\vec{\xi}$
as the vector $\vec q$ is orthogonal to the kernel of $\Lambda$ 
which is spanned by the vector $(1,1,\ldots,1)^T$. In order
to formally solve equation (\ref{eqn:Kxiq}) we can thus use
the Moore-Penrose pseudoinverse of $\Lambda$, which will be
called $T := \Lambda^+$ in the following. Thus we find
\be
    \vec \xi =  \kappa L_{ab} \;   T \vec q.
   \label{eqn:xi-edge}
\ee

The power flow in the perturbed network is to leading order given by
\bea
   F'_{ij} &=& (K_{ij} + \kappa_{ij}) \sin( \phi_i - \phi_j + \xi_i - \xi_j ) \nn \\
    &=& F_{ij}  + \kappa_{ij} \sin( \phi_i - \phi_j)  
         + K_{ij} \cos( \phi_i - \phi_j)   (\xi_i - \xi_j) \nn \\
    &=& F_{ij}  + \kappa_{ij} L_{ij} + \widetilde K_{ij} (\xi_i - \xi_j).
\eea
Inserting the results (\ref{eqn:xi-edge}) for the phase response yields
\be
   F'_{ij} = F_{ij} +  \eta_{ij \leftarrow ab}   \, \kappa,
  \label{eqn:flow-perturbed}
\ee
where we have defined the \emph{link susceptibility}
\bea
\eta_{ij \leftarrow ab} &:=& L_{ab} \times \big[ \widetilde K_{ij} (T_{jb} - T_{ja} - T_{ib} + T_{ia}) \nonumber\\
&\it{} & \quad \quad \quad \quad \quad + (\delta_{ia} \delta_{jb} - \delta_{ja} \delta_{ib}) \big].
\label{eqn:def-eta}
\eea
Related methods are used in power engineering, where they are commonly referred 
to as `line outage distribution factors' \cite{Wood13}. They are used in numerical 
studies of multiple line outages (N-x errors), where a full solution of the all possible 
combinations of contingencies becomes intractable \cite{Gule07}. In this contribution 
we demonstrate the ability of this approach to predict the future dynamics of complex 
supply networks and we provide a detailed network theoretical analysis.

The equations (\ref{eqn:xi-edge}) and (\ref{eqn:flow-perturbed}) 
give the linear response of the phases and the flows to a perturbation
of a single link of the network. Strictly speaking, the are valid only
for infinitesimally small perturbations $\kappa$. It is a  priori
not clear whether these results can be extrapolated to larger
perturbation strength in order to predict the effects of a complete 
breakdown of a single link.

An example of how the damage of a single transmission line
affects the flows in a power grid is shown in Fig.~\ref{fig:flowchange}.
We compare the change of the magnitude of the flow $|F_{ij}'| - |F_{ij}|$ predicted
by linear response theory (\ref{eqn:flow-perturbed})
to the actual value obtained from a numerical solution of the
steady state condition for the original and the perturbed network.
In case of a small damage where only $10 \%$ of the transmission capacity 
is lost ($\kappa_{ab} = - 0.1 \times K_{ab}$) we find an excellent agreement 
between predicted and actual values as expected. 
The situation is more complex in case of a complete breakdown,
i.e.~$\kappa_{ab} = - K_{ab}$. We find that formula (\ref{eqn:flow-perturbed})
reproduces the \emph{relative} changes of the flow surprisingly well.
That is, linear response theory accurately predicts \emph{where} the flow is
rerouted even for a complete breakdown of a transmission line.
However, the magnitude of the flow changes is strongly underestimated
by formula  (\ref{eqn:flow-perturbed}). Consequently, it also fails to reproduce 
that the load of the defective link must vanish, $F_{ab}' \stackrel{!}{=} 0$.
But one can easily fix this deficit by extrapolating the results of linear
response theory if we replace equation $\kappa$ by
$-F_{ab}/\eta_{ab \leftarrow ab}$.
This yields the modified formula
\be
   F_{ij}'' = F_{ij} -  \frac{\eta_{ij \leftarrow ab}} {\eta_{ab \leftarrow ab}} F_{ab} \, .
    \label{eqn:F-linres2}
\ee
In this article, we propose to use this modified linear response formula to
predict impeding overloads which lead to a destabilization of the grid.
An example of a successful application of this method is demonstrated in 
Figure 1 in the main manuscript. 
In the first example, formula  (\ref{eqn:F-linres2}) predicts that
no overload occurs in agreement with the direct solution
of the steady state condition (\ref{eqn:def-steady}). Thus we
expect that the grid relaxes to a new steady state after the
failure of the respective link. This prediction is confirmed by
a direct numerical simulation of the equations of motion.
In the second example  formula  (\ref{eqn:F-linres2}) predicts that
secondary overloads occur after the respective transmission line  failed. 
Indeed, numerical simulations show that
no steady state solution of equation exists
and that the grid becomes unstable and looses synchrony.

\begin{figure}[tb]
\centering
\includegraphics[width=8cm, angle=0]{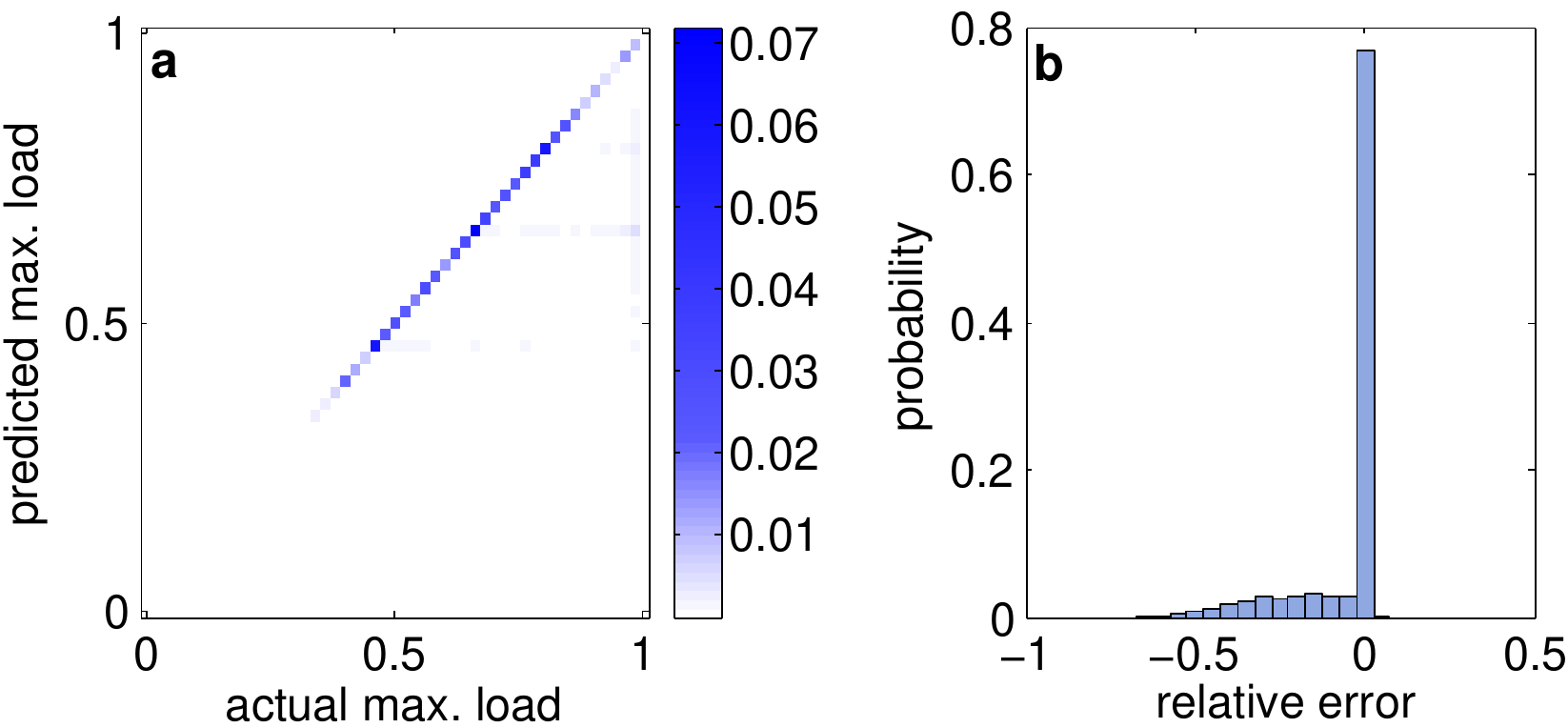}
\caption{\label{fig:lrcomp}
{Quality of load prediction.}
(a) Histogram of the actual maximum load $\ell^{\rm max}_{ab} = \max_{ij}|F_{ij}''/K_{ij}|$
after a the failure of a single link and the prediction based on the linear 
response result (\ref{eqn:F-linres2}). The prediction is very accurate 
in the majority of all cases. Data has been collected for all stable links 
in 400 test networks with the topology of the British grid and 
heterogeneous generation (see section \ref{sec:networkdata} for details).
(b) Histogram of the relative error of the prediction
$(\ell_{\rm predicted}^{\rm max} - \ell_{\rm actual}^{\rm max})/(\ell_{\rm actual}^{\rm max})$.
}
\end{figure}

Furthermore, we test the quality of load prediction 
for stable links in figure \ref{fig:lrcomp}, comparing the predicted maximum 
load after the failure of a single link with the numerically exact values. 
The maximum load $\ell^{\rm max}_{ab} = \max_{ij}|F_{ij}''/K_{ij}|$ 
 is predicted with great accuracy in the majority of all cases.
It is underestimated in approximately 20\% of all cases.

\subsection{Properties of the Laplacian and the susceptibility}
\label{sec:prop-Laplace}

A deeper understanding of the results of the linear response 
approach can be gained using results from graph theory and 
network science. The response $\vec \xi$ is determined by 
the load of the perturbed link and the matrix $\Lambda$ 
through equation (\ref{eqn:Kxiq}). During normal 
operation $\Lambda$  is a Laplacian matrix for which many important
properties are known \cite{Newm10}. These properties are
particularly useful in order to understand how the response
is determined by the structure of the power grid.

\begin{figure}[tb]
\centering
\includegraphics[height=6.5cm, angle=0]{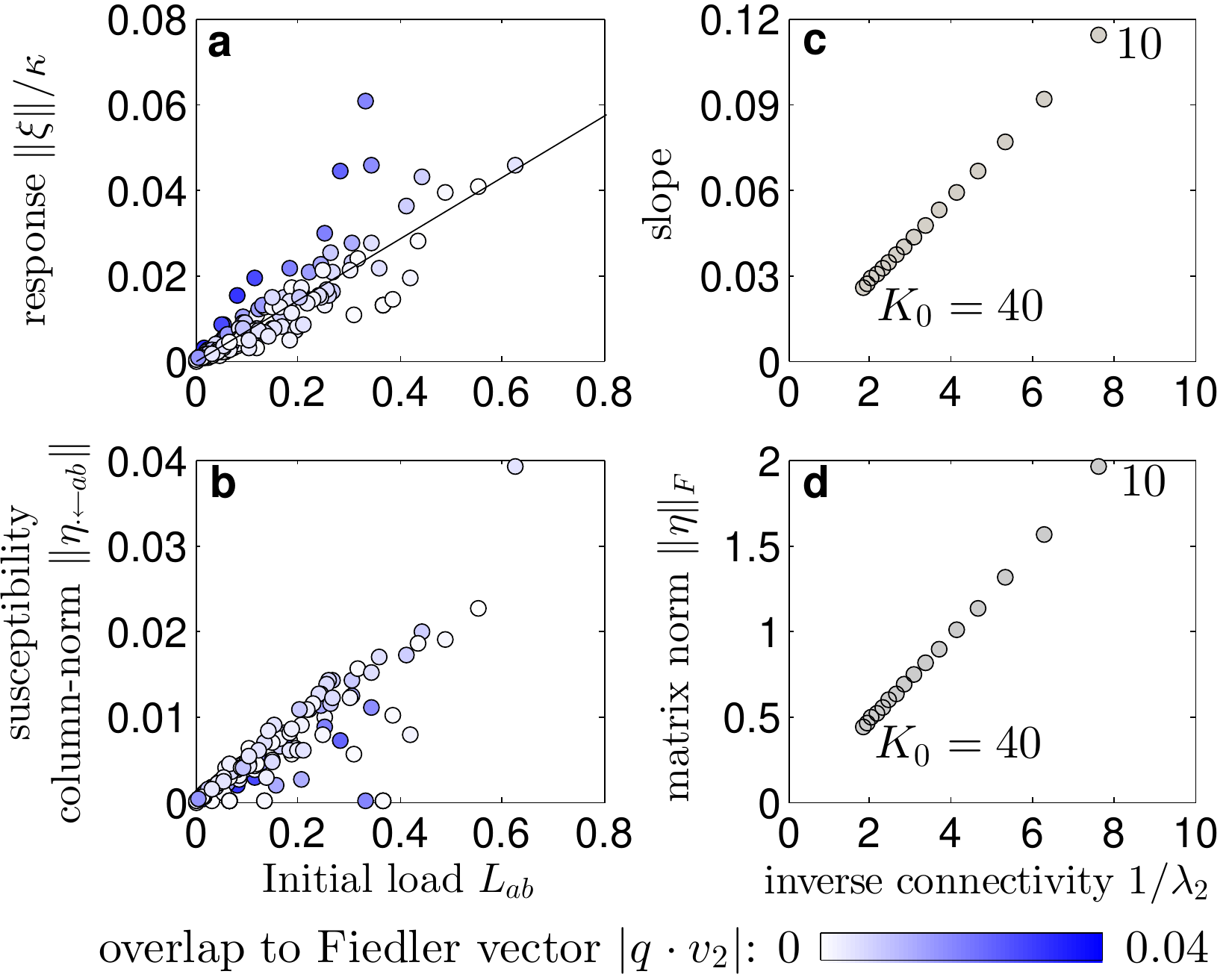}
\caption{\label{fig:sus1}
{The response is proportional to the load of the 
perturbed link and the inverse algebraic connectivity.}
(a,b) Response as a function of the load $L_{ab}$ of the failing 
link. The response of a given network is characterized by the 
norm of the phase difference $\vec{\xi}$ and the columns
of the susceptibility $\eta_{\cdot\leftarrow ab }$.
In addition, the phase difference is significantly enhanced if the 
overlap to the Fiedler vector 
$|\vec{q} \cdot  \vec{v}_2|$ is large, as
indicated by the color code.
(c,d) The global response, averaged over all links 
is proportional to the inverse algebraic connectivity $1/\lambda_2$.
Shown is the slope of the phase response (solid line in (a))
and the matrix norm of the susceptibility $\eta$ for networks with 
different values $K_0=10,12,\ldots,40 \, {\rm s}^{-2}$ and thus 
different connectivity.
The network topology is the same as in Fig.~\ref{fig:flowchange}.
% plotting: aclinks_sus3.m and aclinks_sus4.m 
}
\end{figure}

So assume that $|\phi_i - \phi_j| \le \pi/2$ hold for all links 
of the network. Let $0 = \lambda_1 < \lambda_2 < \cdots \lambda_N$ denote the
eigenvalues of $\Lambda$ and $\vec v_n$ the corresponding 
eigenvectors. Then the response (\ref{eqn:xi-edge}) is
given by
\bea
    \vec  \xi &=& \frac{\kappa F_{ab}}{K_{ab}} \sum_{n=2}^N  \frac{1}{\lambda_n}   
          (\vec v_n \cdot \vec q)  \, \vec v_n.
    \label{eq:response-eigen}      
\eea
The term $n=1$ does not contribute since we have fixed the global
phase such that $\sum \nolimits_j \xi_j = 0$.
This expression shows four important properties:

(1) The response generally scales with the \emph{load} of the defective link
$L_{ab}=F_{ab}/K_{ab}$ times the absolute strength of the perturbation $\kappa$.
This is confirmed by the numerical results for a test grid shown in 
figure \ref{fig:sus1} (a,b).
If we fix the \emph{relative} strength of the perturbation instead,
as for a complete breakdown where $\kappa/K_{ab} = -1$,
then the response scales with the \emph{flow} of the defective link.

(2) The prefactors $1/\lambda_n$ decrease with $n$.
In particular for a heavily loaded network the algebraic connectivity $\lambda_2$
is small such that the term $n=2$
dominates the sum. Then the susceptibility of all links in 
the network scale inversely with the algebraic connectivity of the
network given by the  eigenvalue $\lambda_2$ 
\cite{Fied73,Newm10}. Hence, the response and the susceptibility 
are large if the network defined by the responsive capacities 
$\widetilde K_{ij}$ is weakly connected.

(3) For a heavily loaded network, the link susceptibility scales
with the overlap $|\vec v_2 \cdot \vec q|$, where $\vec v_2$
is the so-called Fiedler vector. This is confirmed by the numerical 
results for a test grid shown in figure \ref{fig:sus1} (a). The 
response is enhanced if the overlap (shown in a color code) 
is large.

The Fiedler overlap can be interpreted
as a measure of the \emph{local algebraic connectivity} of the 
nodes $a$ and $b$. To see this note that the Fiedler vector
can be used to partition a graph into two weakly connected
components \cite{Newm10,Fort10}. The overlap with the 
vector $\vec q$ is largest if the two nodes $a$ and $b$ are 
in different components and thus weakly connected.

To illustrate this, let us consider the limiting case 
$\lambda_2 \rightarrow 0$, i.e. the case where the network
defined by the responsive capacities $\widetilde K_{ij}$ becomes disconnected 
into two fragments with $N_1$ and $N_2$ nodes, respectively.
The Fiedler vector then reads 
\be
   \vec v_2 = \frac{1}{\sqrt{N_1+N_2}}  \;
     ( \underbrace{ \sqrt{N_2/N_1},\ldots}_{N_1 \; \mbox{times}} \, ,
      \underbrace{-\sqrt{N_1/N_2} ,\ldots}_{N_2 \; \mbox{times}}  )^T , 
\ee
it is positive for the nodes in the first fragment and negative for the 
nodes in the other fragment. The overlap is then given by  
\be
    |\vec v_2 \cdot \vec q| =  \left\{
    \begin{array}{l c l}  
      0  & \,  \mbox{if  $a$ and $b$ are in the same fragment} \\  
      \sqrt{\frac{N_1+N_2}{N_1 N_2}}  & \,  
         \mbox{if $a$ and $b$ are in different fragments.} \\  
   \end{array} \right.  \nn
\ee

(4) In the limit of a disconnected network  the response $\vec \xi$ to
a perturbation at the link $(a,b)$ diverges if it links the
weakly-connected components. If the perturbation occurs within
one component, then the response remains finite.

(5) On a global scale the response $\vec \xi$ and the susceptibility $\eta$ are essentially
given by the inverse algebraic connectivity $1/\lambda_2$ of the network.
We have already discussed the importance of the term $n=2$ for
the response (\ref{eq:response-eigen}). Averaging over all possible 
trigger links, this term dominates even for a strongly coupled 
network.
To test this proposition, we have calculated $\vec \xi$ and $\eta$ 
for a test grid with different values of the global coupling strength, 
i.e.~we write $K_{ij} = K_0 A_{ij}$ and vary the global prefactor $K_0$.
The numerical results shown in figure \ref{fig:sus1} (c,d) confirm 
that the norm of the response $\| \vec \xi \|$ averaged over all trigger links
as well as the Frobenius norm of the susceptibility matrix $\eta$
are almost perfectly proportional to $1/\lambda_2$.

\section{Alternative identification statistics and additional results}

\subsection{Introducing novel indicator variables}

In this contribution we introduce three novel indicator variables for distinguishing 
critical and stable links on the basis of the topology of the network and its loads 
\emph{prior to} a line failure. All three approaches are based on the fact that network 
flow must be rerouted when a transmission line fails. We expect that this is impossible 
when the \emph{ratio of flow and redundant capacity}
\begin{equation}
  r_{ab} = |F_{ab}/K_{ab}^{\rm red}| \label{eqn:def-pred1}
\end{equation}
as defined in \ref{sec:kred} is too large. Alternatively, we expect a global breakdown, 
when linear response theory as introduced in section \ref{sec:linres} predicts secondary 
overloads, i.e. when the \emph{predicted maximum load}
\begin{equation}
  \ell_{ab}^{\rm max} = \max_{(i,j)} |F''_{ij}/K_{ij}| \label{eqn:def-pred2}
\end{equation}
is too large. Additionally we test, whether a combination of both variables
\begin{equation}
 c_{ab} = \sqrt{ r_{ab}^2 + \ell_{ab}^{{\rm max} \, 2}   } \,
    \label{eq:combined-pre}
\end{equation}
can be an even more successful indicator. We compare the performance of these 
three novel indicator variables with other indicators for the importance 
of links in power-grids, namely the load $L_{ab} = |F_{ab}/K_{ab}|$, 
the flow $|F_{ab}|$, 
the topological edge-connectivity $\tau_{ab}$ and
the edge betweenness centrality $\epsilon_{ab}$ \cite{Newm10}.

%\footnote{We use the symbol $\tau$ for the topological edge-connectivity instead
%of the more common symbol $\lambda$, as this is already used for the algebraic
%connectivity.}

Prediction or classification success does not only depend on the information comprised 
in an indicator variable, but also on the way, this information is used. In this contribution 
we test all indicator variables using two different methods of determining decision margins, 
i.e. values which announce critical links: thresholding and a naive Bayesian classifier 
\cite{Rish01, Zhan04}. Thresholding is a very robust approach, since it does not depend 
on the amount of training data or training method. Additionally, all considerations 
concerning the rerouting of network flow support the idea that large values of the 
indicators $r_{ab}$, $\ell_{ab}^{\max}$ and $c_{ab}$ are linked to the criticality of links.
Therefore predictions of critical links can be made by simply imposing a threshold $h$ which 
distinguishes between indicator values related to critical links or stable links.
On the other hand, a naive Bayesian classifier does not require any assumption on the 
system that generated the data set or the range of suitable indicator values.
Additionally it is able to detect relations between indicator variable and target event 
that are more complex than a linear correlations.

\subsection{Quantifying prediction success}

The success of indicators and classifiers is typically evaluated using contingency tables. 
A contingency tables (see Tab.~\ref{tab:example} for an example) summarizes 
the numbers of occurrences $n_{\rm TP}$, $n_{\rm FP}$, $n_{\rm FN}$ and $n_{\rm TN}$ 
of four possible outcomes of a prediction task:\\

\begin{tabular}{ll}
  \emph{True positive} (TP): & Link is predicted critical\\
  & and is critical.\\
  \emph{False positive} (FP): & Link is predicted critical \\
  & but is stable.\\
  \emph{False negative} (FN): & Link is predicted stable \\
  & but is critical.\\
  \emph{True negative} (TN): & Link is predicted stable \\
  & and is stable.\\
\end{tabular}\\

\begin{table}[ht]
\centering
\begin{tabular}{| c || c | c |} \hline
\hfill Link is & critical & stable  \\ \hline \hline
predicted & \multirow{2}{*}{$n_{\rm TP}$} & \multirow{2}{*}{$n_{\rm FP}$}\\
critical &  &  \\ \hline
predicted & \multirow{2}{*}{$n_{\rm FN}$} & \multirow{2}{*}{$n_{\rm TN}$}\\
stable & & \\ \hline \hline
& \multirow{2}{*}{$n_{\rm TP} + n_{\rm FN} =n_{\mbox{critical}}$}&  \multirow{2}{*}{$n_{\rm FP} + n_{\rm TN} = n_{\mbox{stable}}$}\\
&  & \\
\hline
\end{tabular}
\caption{\label{tab:example}Example for a contingency table}
\end{table}

\begin{figure}
  \includegraphics[width=8cm, trim=30mm 22mm 30mm 22mm]{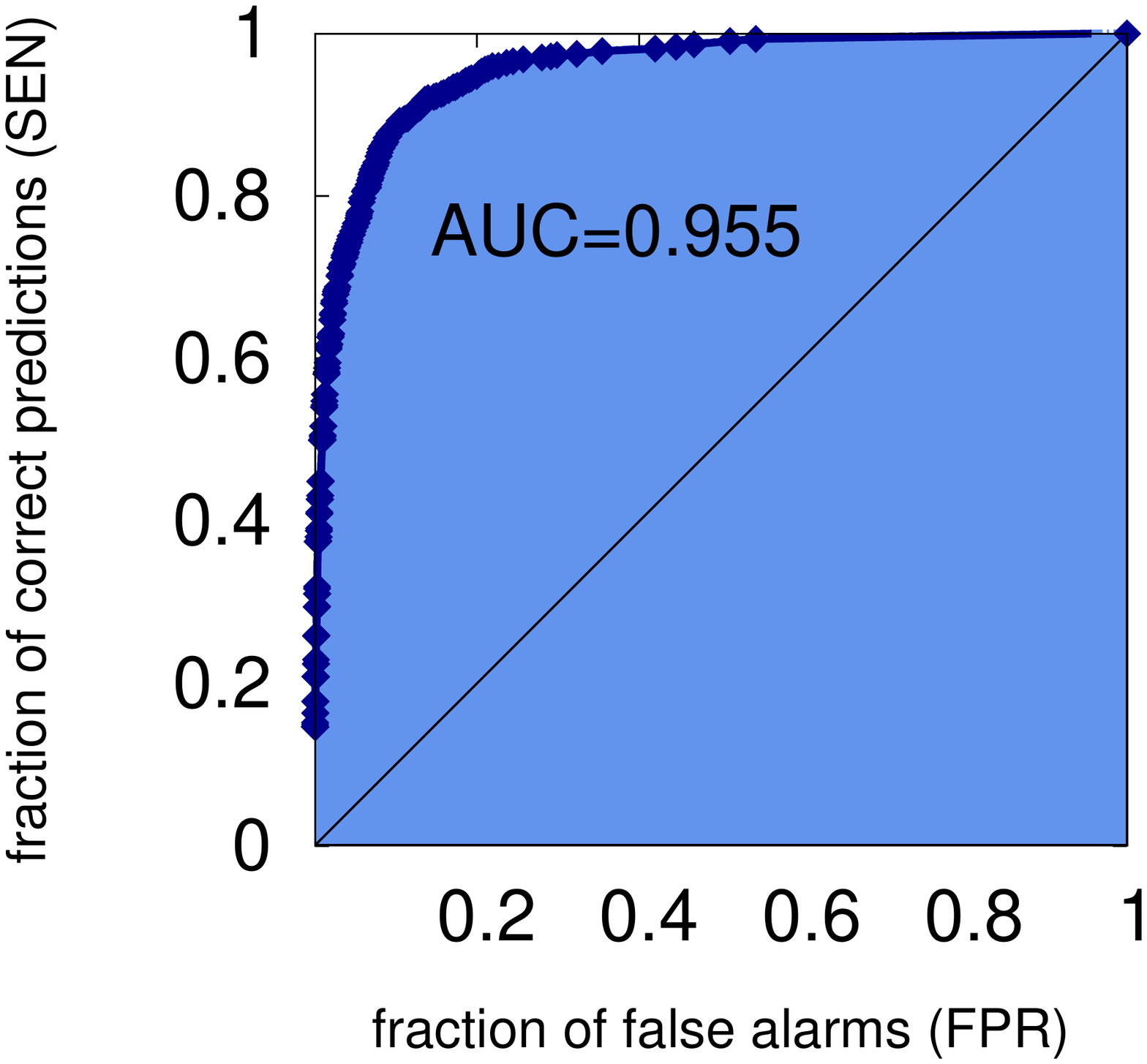}
  \caption{\label{auc-example} 
    {An example for the calculation of an ROC-curve and 
    the area under curve (AUC).}
  }
\end{figure}

A good indicator aims on maximizing $n_{\rm TP}$ while simultaneously minimizing $n_{\rm FP}$.
Setting these quantities in relation to the total numbers of critical and stable links, $n_{\rm critical}$ 
and $n_{\rm stable}$, one can generate receiver operating characteristics (ROC). In more detail, 
the fraction of correct predictions, also called the true positive rate or the sensitivity
\be
  \mbox{SEN} := \frac{n_{\rm TP}}{n_{\rm critical}} 
  \label{eqn:def-sen}
\ee
is compared to the fraction of false alarms, also called the false positive rate
\be
   \mbox{FPR} := \frac{n_{\rm FP}}{n_{\rm stable}}.
\ee
If the classification depends on a threshold value $h$ one can plot
$\mbox{SEN}$ vs.~$\mbox{FPR}$ for different values of $h$ to obtain a 
ROC-curve (see Fig.~\ref{auc-example} for an example). Making no 
predictions and therefore also no false predictions corresponds to the 
point $(\mbox{FPR},\mbox{SEN}) = (0,0)$. Classifying every link to be 
critical generates the point $(\mbox{FPR},\mbox{SEN}) = (1,1)$.
An ROC-curve made by random predictions consists in a straight line with
slope $1$ through the origin. A perfect indicator creates an ROC-curve 
that contains the point $(\mbox{FPR},\mbox{SEN}) =  (0,1)$.
Consequently, a classifier is judged to be the better, the nearer the 
ROC-curve approaches the the point $(0,1)$, i.e. the upper left corner of the plot.

Typically, \emph{summary indices} such as the area under curve (AUC) 
(see Figs.~\ref{auc-example} and \ref{auc-lin-bayes-comp}) or the 
distance to the upper left corner of the ROC plot are used to compare 
ROC curves of several different indicators.

\subsection{Thresholding approach for determining decision margins}
\label{sec:linclass}

\begin{figure}[tb]
\centerline{
  \includegraphics[width=8cm]{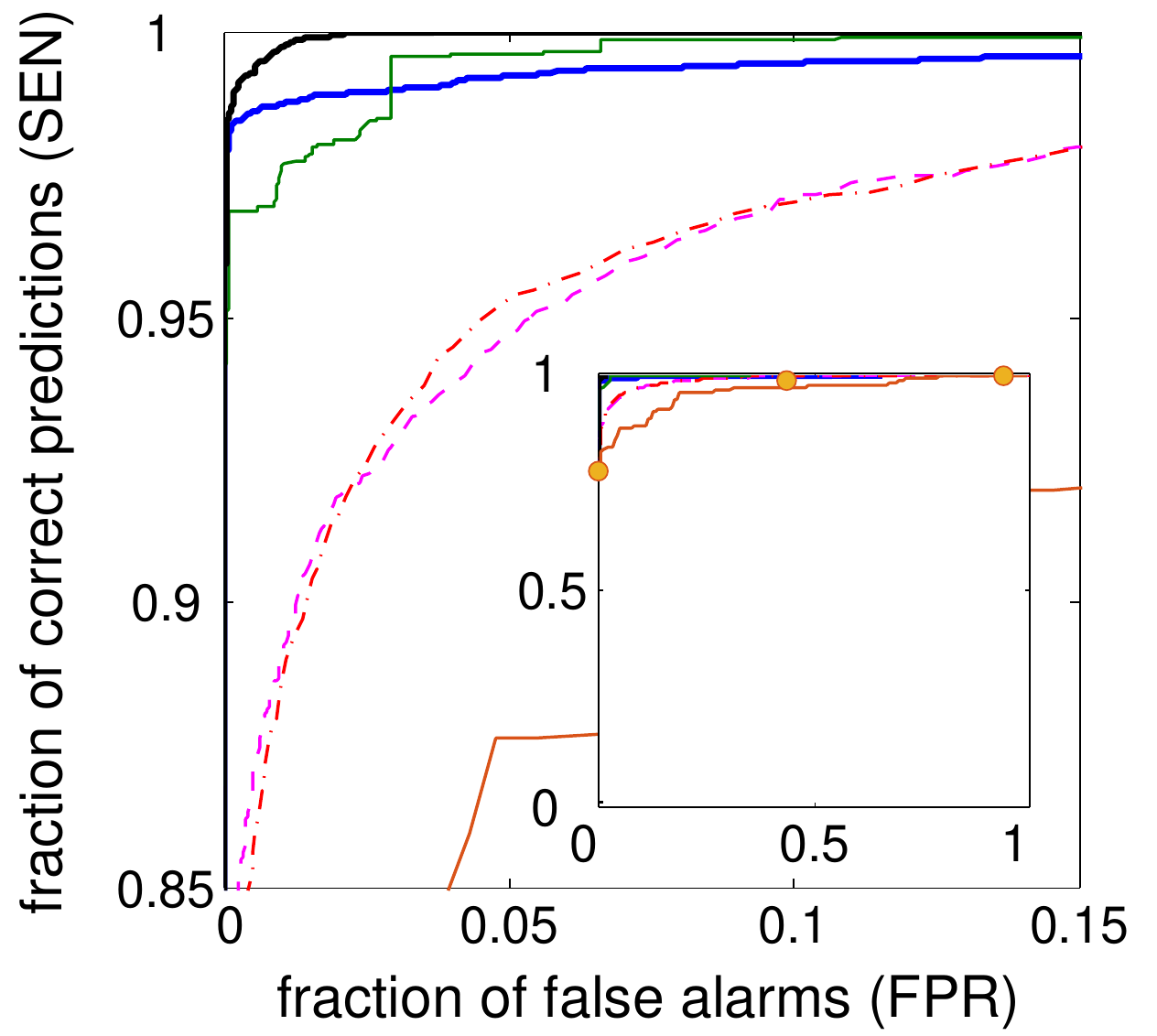}
  }
\caption{
\label{fig:roc1}
{Receiver operating characteristic curves of different indicators for critical and stable links.}
The ROC curve shows the fraction of correct predictions of the 
classifier vs.~the fraction of false alarms for various values of the threshold 
value $h$. The proposed classifiers $r_{ab}$ (\textcolor{blue}{\protect\rule[0.5ex]{1.2em}{1pt}}), 
$\ell_{ab}^{\rm max}$ (\textcolor{green}{\textbf{---}}) and $c_{ab}$ 
(\protect\rule[0.5ex]{1.2em}{1pt}) closely approach the perfect value 
$(\mbox{FPR},\mbox{SEN}) =  (0,1)$.
They clearly outperform classification schemes based on the local 
measures load $L_{ab}$ (\textcolor{magenta}{$- - -$}) 
and flow $|F_{ab}|$ (\textcolor{red}{$- \cdot -$}).
Purely topological quantites such as the the 
topological edge-connectivity $\tau_{ab}$ ($\circ$)
and the edge betweenness centrality $\epsilon_{ab}$ (\textcolor{brown}{\textbf{---}})
perform even worse.
Note that the ROC curve for $\lambda_{ab}$ consists of discrere
points only as the variable $\lambda_{ab}$ itself is discrete. 
The inset shows the entire ROC curves, while the main panel
shows a magnification around the point $(\mbox{FPR},\mbox{SEN}) =  (0,1)$.
The curves are based on the simulation of 400 random realizations 
of the British power grid with heterogeneous generation  
(cf.~section \ref{sec:networkdata}).
}
\end{figure}

All considerations concerning the rerouting of network flow support the idea that 
large values of the indicators $r_{ab}$, $\ell_{ab}^{\rm max}$ and $c_{ab}$ occur, 
if the corresponding links are critical. Assuming the relation between an indicator 
variable $x_{ab}$ and the tendency of an link to be critical is linear, we can use a 
simple discrimination threshold $h$ to separate values that predict critical links from others:
\bea
   &&  x_{ab}  > h  \;  \Rightarrow \;  
                       \mbox{predicted to be critical}, \nn \\
   &&  x_{ab}  \le h  \; \Rightarrow \;  
                    \mbox{predicted to be stable}.
   \label{eqn:advclass}
\eea
More precisely, we propose two classification systems:
(a) Based on the quantification of the residual capacity defined
in section \ref{sec:kred} we propose the following classification system:
\bea
   &&  r_{ab}  > h   \;  \Rightarrow \;  
                       \mbox{predicted to be critical}, \nn \\
   &&  r_{ab}  \le h \; \Rightarrow \;  
                    \mbox{predicted to be stable}, 
   \label{eqn:def-class}
\eea
where $h$ is a threshold value that can be optimized for the
specific task.
(b) Based on the results of linear response theory introduced in
section \ref{sec:linres} we propose the following classification system:
\bea
   && \ell_{ab}^{\rm max} >    h   \;  \Rightarrow \;  \mbox{predicted to be critical}, \nn \\
   && \ell_{ab}^{\rm max} \le h   \; \Rightarrow \;  \mbox{predicted to be stable}.
   \label{eqn:def-class2}
\eea
Bridges are always predicted to be critical. The variable $h$ denotes
a threshold value that can be optimized for a specific task. 

In the current setting, the number of false positive predictions 
can be minimized by choosing a high value of $h$, while the number
of false negative predictions can be minimized by choosing 
a small value of $h$.

\begin{table}[tb]
\centering
\begin{tabular}{| r | r | r | r | r |} \hline
(a) \hfill Link is & critical & stable & \\ \hline
$r_{ab}  > h $   &   \cellcolor{lgreen} 8650 &  \cellcolor{lred}         279   & $\mbox{PPV}=96.88  \%$   \\ \hline
$r_{ab}  \le h $    &   \cellcolor{lred}         116 &  \cellcolor{lgreen}  56955  & $\mbox{NPV}=99.80 \%$   \\ \hline
& $\mbox{SEN}=98.68  \%$  &  $\mbox{SPE}=99.51 \%$ & \\ \hline
\end{tabular}
\phantom{0}
\begin{tabular}{| r | r | r | r | r |} \hline
(b) \hfill Link is & critical & stable & \\ \hline
$r_{ab} > h $    &   \cellcolor{lgreen}  8433 &  \cellcolor{lred}          0   & $\mbox{PPV}=100.0  \%$   \\ \hline
$r_{ab} \le h $     &   \cellcolor{lred}         333 &  \cellcolor{lgreen}  57234  & $\mbox{NPV}=99.42 \%$   \\ \hline
& $\mbox{SEN}=96.20  \%$  &  $\mbox{SPE}=100.0 \%$ & \\ \hline
\end{tabular}
\phantom{0}
\begin{tabular}{| r | r | r | r | r |} \hline
(c) \hfill Link is & critical & stable & \\ \hline
$r_{ab}  > h $    &   \cellcolor{lgreen}  8766 &  \cellcolor{lred}    37574   & $\mbox{PPV}=18.92  \%$   \\ \hline
$r_{ab}  \le h $    &   \cellcolor{lred}        0 &  \cellcolor{lgreen}  19660  & $\mbox{NPV}=100.0 \%$   \\ \hline
& $\mbox{SEN}=100.0  \%$  &  $\mbox{SPE}=34.35 \%$ & \\ \hline
\end{tabular}
\caption{
\label{tab:ct-gbhetc2-cap}
Performance of a classification system of critical and stable links 
according to the redundant capacity as defined in Eq.~(\ref{eqn:def-class}).
Contingency table for the three different values of the threshold $h$
optimized for different tasks: 
(a) Threshold value $h=0.614$ for which the ROC curve approaches the
perfect operation point $(0,1)$ most closely.
(b) No false positive results occur for high threshold value $h=0.84$.
(c) No false negative results occur for a low threshold value $h=0.0493$.
Results are summarized for 400 random realizations of the british power grid 
with heterogeneous generation (cf.~section \ref{sec:networkdata}).
}
\end{table}

The resulting ROC curves for the identification of critical and
stable links are shown in Fig.~\ref{fig:roc1}.
The proposed classifier  based on the redundant capacity
(\ref{eqn:def-class}) and  based on linear response theory
(\ref{eqn:def-class2}) closely approach the perfect value 
$(\mbox{FPR},\mbox{SEN}) =  (0,1)$. 
They clearly outperform classification schemes based on the local 
measures load $L_{ab}$ and flow $|F_{ab}|$. Purely topological quantities 
such as the the topological edge-connectivity $\tau_{ab}$ and the edge
betweenness centrality $\epsilon_{ab}$ perform even worse 
(cf.~also \cite{Hine10}).

\begin{table}[tb]
\centering
\begin{tabular}{| r | r | r | r | r |} \hline
(a) \hfill Link is & critical & stable & \\ \hline
  $\ell_{ab}^{\rm max} >  h$ &   \cellcolor{lgreen}  8554 &  \cellcolor{lred} 708   & $\mbox{PPV}=92.36 \%$   \\
  or bridge & & & \\ \hline
$\ell_{ab}^{\rm max} \le  h$    &   \cellcolor{lred}         212 &  \cellcolor{lgreen}  56526  & $\mbox{NPV}=99.63 \%$   \\ \hline
& $\mbox{SEN}=97.58 \%$  &  $\mbox{SPE}=98.76 \%$ & \\ \hline
\end{tabular}
\phantom{0}
\begin{tabular}{| r | r | r | r | r |} \hline
(b) \hfill Link is & critical & stable & \\ \hline
  $\ell_{ab}^{\rm max} >  h$ &   \cellcolor{lgreen}  8238 &  \cellcolor{lred}          0   & $\mbox{PPV}=100.0  \%$   \\
 or bridge & & & \\ \hline
$\ell_{ab}^{\rm max} \le  h$      &   \cellcolor{lred}    528 &  \cellcolor{lgreen}  57234  & $\mbox{NPV}=99.09 \%$   \\ \hline
& $\mbox{SEN}=93.98 \%$  &  $\mbox{SPE}=100.0 \%$ & \\ \hline
\end{tabular}
\phantom{0}
\begin{tabular}{| r | r | r | r | r |} \hline
(c) \hfill Link is & critical & stable & \\ \hline
  $\ell_{ab}^{\rm max} >  h$ &   \cellcolor{lgreen}  8766  &  \cellcolor{lred}  19866 & $\mbox{PPV}=30.62  \%$   \\
  or bridge & & & \\ \hline
$\ell_{ab}^{\rm max} \le  h$     &   \cellcolor{lred}        0 &  \cellcolor{lgreen}  37368  & $\mbox{NPV}=100.0 \%$   \\ \hline
& $\mbox{SEN}=100.0 \%$  &  $\mbox{SPE}=65.29 \%$ & \\ \hline
\end{tabular}
\caption{
\label{tab:ct-gbhetc2-lin}
Performance of a classification system of critical and stable links 
based on linear response theory as defined in Eq.~(\ref{eqn:def-class2}).
Contingency table for the three different values of the threshold $h$
optimized for different tasks: 
(a) Threshold value $h=0.977$ for which the ROC curve approaches the
perfect operation point $(0,1)$ most closely.
(b) No false positive results occur for a high threshold value $h=1.066$.
(c) No false negative results occur for a low threshold value $h=0.733$.
Results are summarized for 400 random realizations of the British power grid 
with heterogeneous generation (cf.~section \ref{sec:networkdata}).
}
\end{table}

\begin{table}[tb]
\centering
\begin{tabular}{| r | r | r | r | r |} \hline
(a) \hfill Link is & critical & stable & \\ \hline
  $c_{ab} >  h$ &   \cellcolor{lgreen}  8723 &  \cellcolor{lred}         336   & $\mbox{PPV}=96.29 \%$   \\
  or bridge & & & \\\hline
$c_{ab} \le  h$    &   \cellcolor{lred}  43 &  \cellcolor{lgreen}  56898  & $\mbox{NPV}=99.92 \%$   \\ \hline
& $\mbox{SEN}= 99.51 \%$  &  $\mbox{SPE}= 99.41 \%$ & \\ \hline
\end{tabular}
\phantom{0}
\begin{tabular}{| r | r | r | r | r |} \hline
(b) \hfill Link is & critical & stable & \\ \hline
  $c_{ab} >  h$ &   \cellcolor{lgreen}  8472 &  \cellcolor{lred}          0   & $\mbox{PPV}=100.0  \%$   \\
  or bridge & & & \\ \hline
$c_{ab} \le  h$      &   \cellcolor{lred}  294 &  \cellcolor{lgreen}  57234  & $\mbox{NPV}=99.49 \%$   \\ \hline
& $\mbox{SEN}=96.65  \%$  &  $\mbox{SPE}=100.0 \%$ & \\ \hline
\end{tabular}
\phantom{0}
\begin{tabular}{| r | r | r | r | r |} \hline
(c) \hfill Link is & critical & stable & \\ \hline
  $c_{ab} >  h$ &   \cellcolor{lgreen}  8766  &  \cellcolor{lred}  1198 & $\mbox{PPV}=87.98  \%$   \\
  or bridge & & & \\ \hline
$c_{ab} \le  h$     &   \cellcolor{lred}        0 &  \cellcolor{lgreen}  56036  & $\mbox{NPV}=100.0 \%$   \\ \hline
& $\mbox{SEN}=100.0  \%$  &  $\mbox{SPE}=97.91 \%$ & \\ \hline
\end{tabular}
\caption{
\label{tab:ct-combined}
Performance of a classification system of critical and stable links 
based on the combined indicator $c_{ab}$ as defined in Eq.~(\ref{eqn:def-class3}).
Contingency table for the three different values of the threshold $h$
optimized for different tasks: 
(a) Threshold value $h=1.044$ for which the ROC curve approaches the
perfect operation point $(0,1)$ most closely.
(b) No false positive results occur for a high threshold value $h=1.289$.
(c) No false negative results occur for a low threshold value $h=0.99$.
Results are summarized for 400 random realizations of the British power grid 
with heterogeneous generation (cf.~section \ref{sec:networkdata}).
}
\end{table}

Finally, the threshold value $h$ can be chosen according to the specific
application of the classification system. A common choice is the point
which has the smallest distance to the perfect operational point
$(\mbox{FPR},\mbox{SEN}) =  (0,1)$.
This choice represents a compromise between 
the conflicting goals of large sensitivity an small false positive rate. 
For the present network data set it yields the value
$h=0.614$ for the classification system (\ref{eqn:def-class}) and
$h=0.977$ for the classification system (\ref{eqn:def-class2}).
In these cases, the absolute fraction of false predictions lies below
$0.75 \%$ or $1.39 \%$, respectively (cf.~Tables \ref{tab:ct-gbhetc2-cap} (a) 
and \ref{tab:ct-gbhetc2-lin} (a)).
A different strategy would be to minimize the number of false 
positive results.  For the given network data this number can 
be reduced to zero by choosing a high value of $h$
(cf.~Tables \ref{tab:ct-gbhetc2-cap} (b) and 
\ref{tab:ct-gbhetc2-lin} (b)). Notably, the sensitivity still remains 
reasonably large in this case.
Similarly, the number of false negative results can be reduced to 
zero by chosing a small threshold value 
(cf.~Tables \ref{tab:ct-gbhetc2-cap} (c) and \ref{tab:ct-gbhetc2-lin} (c)).
This strategy detects all critical links but also  yields a 
comparably large number of false positive predictions.

The results are summarized in the contingency tables 
\ref{tab:ct-gbhetc2-cap} and \ref{tab:ct-gbhetc2-lin} .
In addition to the absolute number of true and false predictions, we also
give some statistical quantities to characterize the classifier:
the sensitivity (\ref{eqn:def-sen}), the specificity 
$\mbox{SPE} := 1 - \mbox{FPR}$ as well as the
positive predictive value
\be
    \mbox{PPV} := \frac{n_{\rm TP}}{n_{\rm TP} + n_{\rm FP}}  \nn
  \label{eqn:def-ppv}
\ee
and the negative predictive value
\be
   \mbox{NPV} := \frac{n_{\rm NP}}{n_{\rm TP} + n_{\rm FN}}  \, .\nn
\ee

Combining the redundant capacity and the linear response theory according to 
Eq.~(\ref{eq:combined-pre}) yields a combined indicator
\begin{eqnarray}
 && c_{ab} >    h   \;  \Rightarrow \;  \mbox{predicted to be critical}, \nn \\
   && c_{ab} \le h   \; \Rightarrow \;  \mbox{predicted to be stable}. 
   \label{eqn:def-class3}
\end{eqnarray}
As shown by the ROC curve in Fig.~\ref{fig:roc1} the combined indicator further 
improves the performance. Most importantly, we can achieve a sensitivity of 100 \% 
for the current dataset by choosing a threshold $h \le 0.99021$. For this choice 
the false positive rate is still below $2.1 \%$. Using a threshold $h=1.28$ reduces 
the false positive rate to zero at a sensitivity of $96.65 \%$.

\subsection{A Bayesian approach for classifying critical links}
\label{sec:adv_stat}

In the previous Sec.~\ref{sec:linclass}, values of indicator variables are selected by 
imposing a threshold. This simple and robust approach is sufficient, if the relationship 
between the values of the indicator variable and the prediction success is monotonous.
A more general approach that does not require any knowledge about the system 
under study consists in classifying suitable values of indicator variables through 
conditional probability density functions (CPDFs), i.e., by using naive empirical Bayesian 
classifiers \cite{Rish01, Zhan04}. The resulting ROC curves are \emph{proper} 
ROC curves, i.e.~, they are concave and consequently have a monotonously 
decreasing slope \cite{Egan75}. This effect is obtained by choosing relevant 
indicator values according to the probability that the link $(a,b)$ is critical 
given a indicator value $x_{ab}$. To express this in formulas, we introduce the
 binary event variable 
\begin{equation}
   \chi = \left\{ \begin{array}{ll}
          1 & \mbox{: if } (a,b) \; \mbox{is critical};\\
          0 & \mbox{: if } (a,b) \; \mbox{is stable}.
          \end{array}\right.
\end{equation}
Decision making, i.e.~here predicting criticality of the link is then done using a 
probability threshold $h'$
\bea
   &&  p(\chi = 1|x_{ab})  \ge h'  \;  \Rightarrow \;  
                       \mbox{predicted to be critical}, \nn \\
   &&  p(\chi = 1|x_{ab})  < h'  \; \Rightarrow \;  
                    \mbox{predicted to be stable},
   \label{eqn:bayesclass}
\eea
with $x_{ab}$ denoting the value of a indicator variable computed for the 
link $(a,b)$. Changing $h'$ from the global maximum of the CPDF 
$p(\chi = 1|x_{ab})$ to $0$, generates an ROC curve starting from the 
lower left corner to the upper right. 

Although this approach is mathematically well defined, there are several (technical) issues 
that have an influence on the resulting ROC curves. In general one observes an increase 
in the area under curve (AUC), with increasing number of bins used to estimated the 
CPDF from a given data set.
However, allowing an arbitrarily large number of bins leads to ROCs which are specific for 
the data set under study (over-fitting) and are irreproducible when using other data sets 
generated by the same system under study. Therefore we determine the number of bins 
in an adaptive way, i.e. by choosing it such that each bin of the CPDF has at least two entries.
Depending on the range of indicator variables it can be more appropriate to use the logarithm 
of a indicator variable and not the variable itself to determine suitable binnings.

\begin{figure}[t!!!]
    \centerline{
    \includegraphics[angle=0, width=8cm]{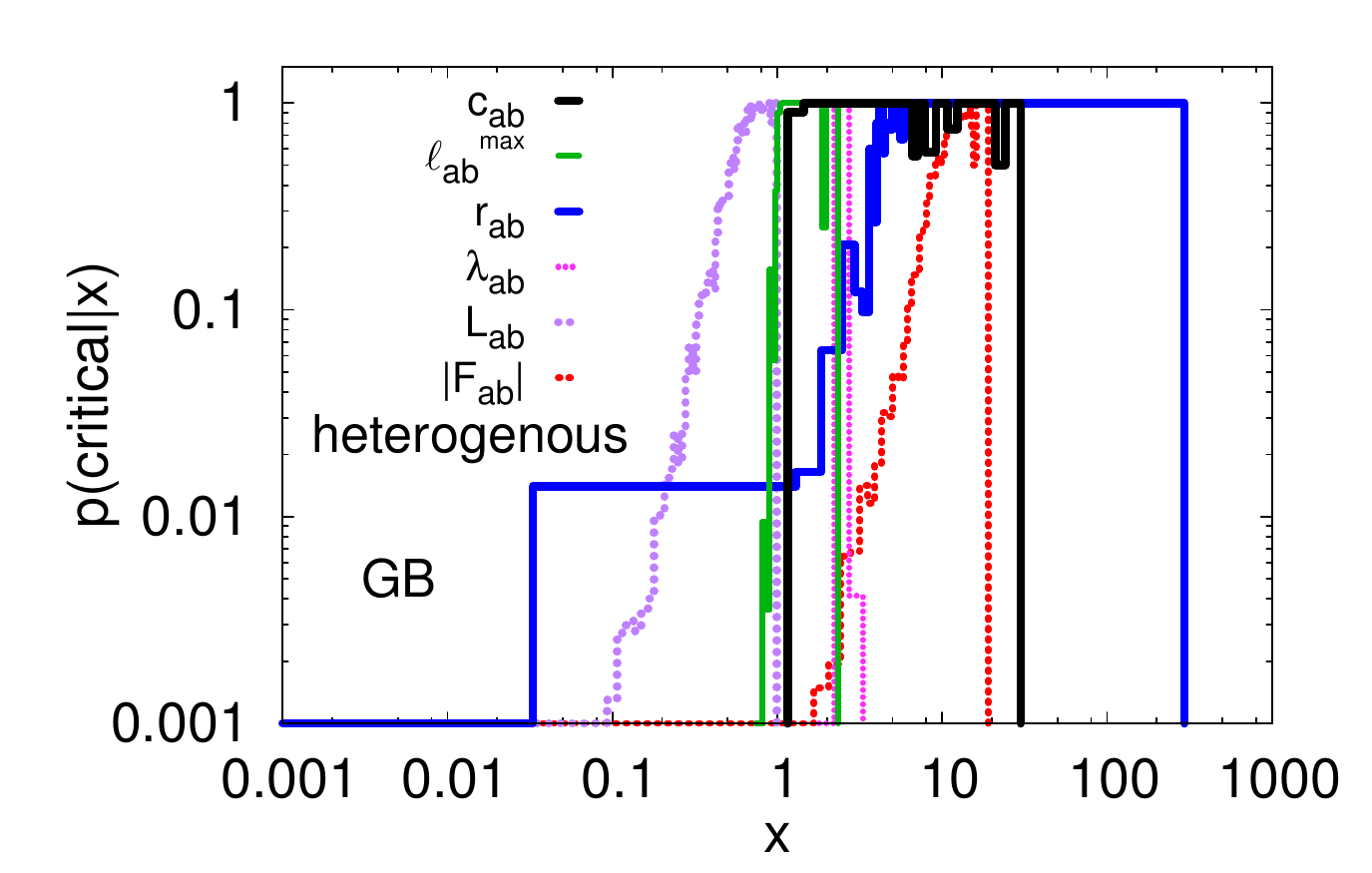}
  }
\caption{\label{cpdf-example} 
Cumulative probability distribution function (CPDF) for all indicator variables calculated for 
the British grid and heterogeneous power generation. In order to cover the wide range of 
indicator variables, the binning was done using irregular bin sizes on logarithmic scales, 
if appropriate. Number and size of bins were adapted such that each bin is based on at least two counts.
}
\end{figure}

\begin{figure}[tb]
  \centerline{
   \includegraphics[width=4cm]{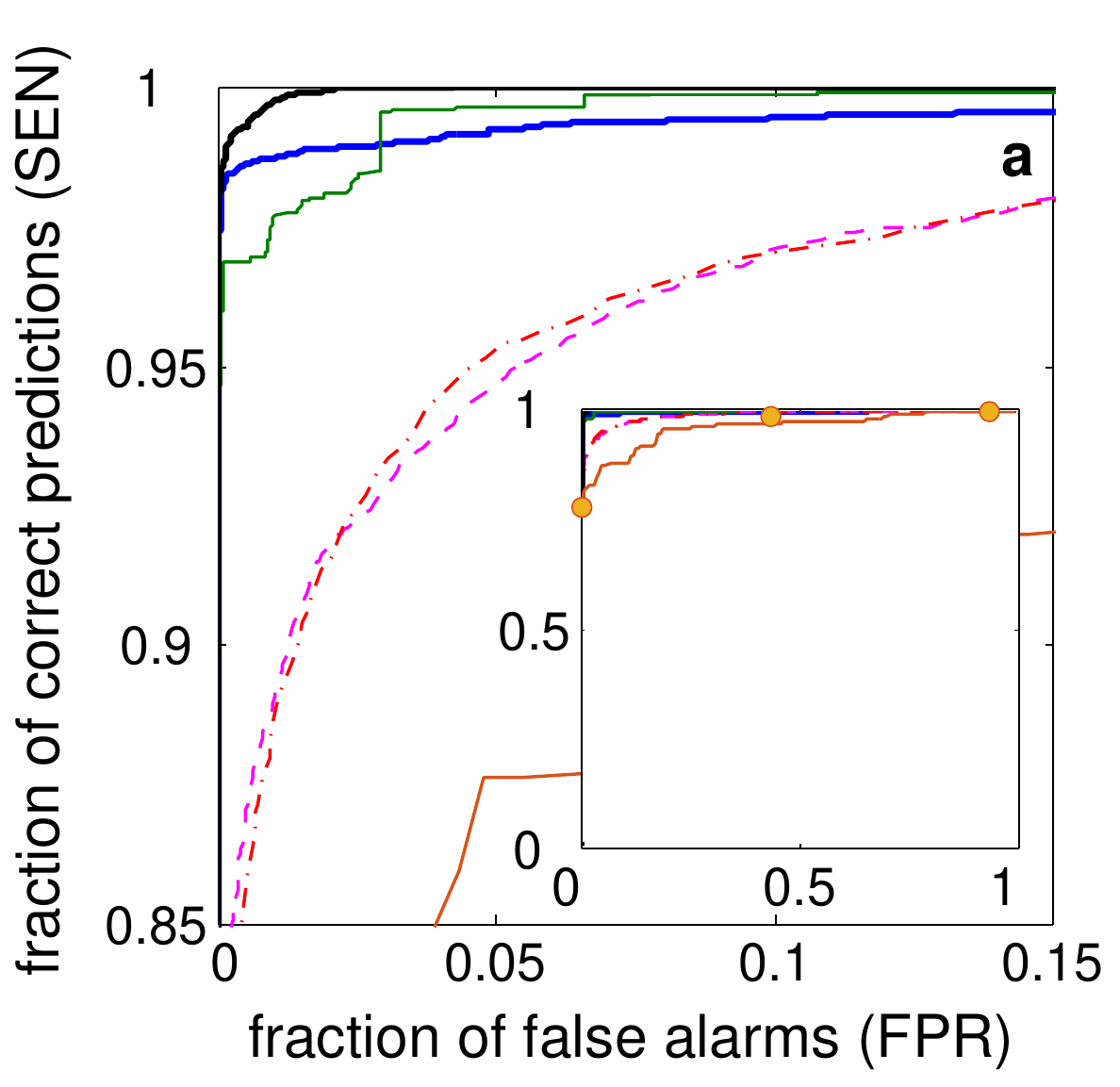}
   \hspace*{0.1cm}
   \includegraphics[width=4cm]{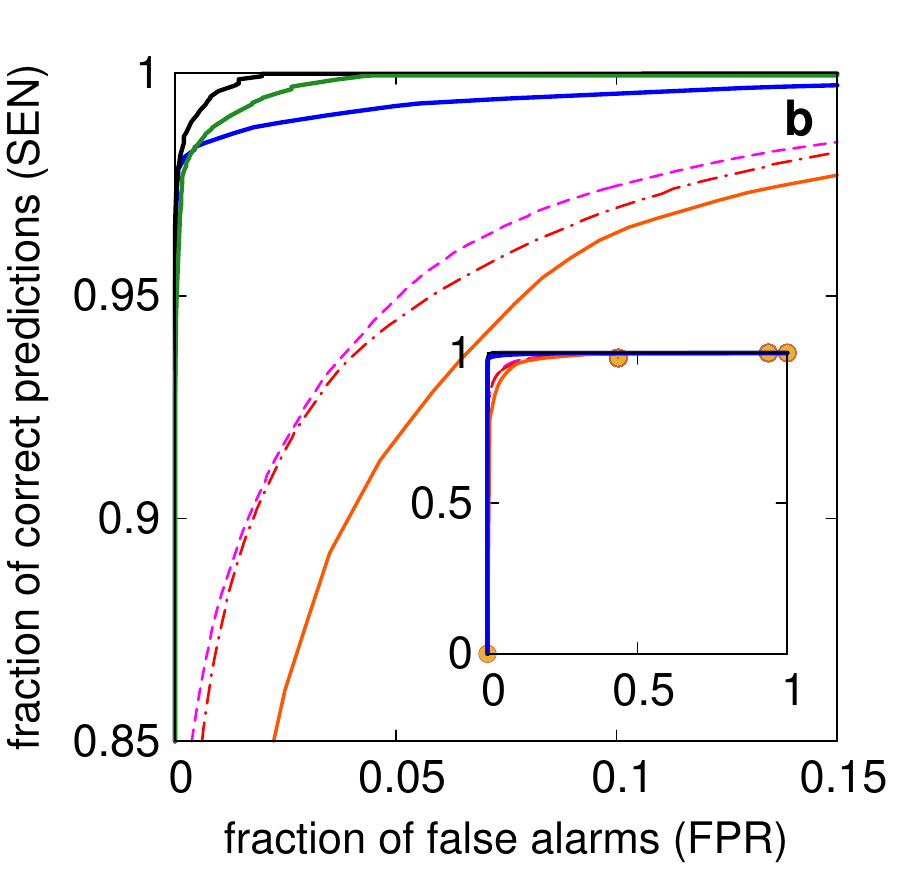}
   }
\caption{\label{roc-lin-bayes-comp}
{ROC curves of different classifiers for critical links using (a) the linear classifier and 
(b) the Bayesian prediction scheme:} 
the ratio $r_{ab}$ (\textcolor{blue}{\protect\rule[0.5ex]{1.2em}{1pt}}),
the predicted max. load $\ell_{ab}^{\rm max}$  (\textcolor{green}{\textbf{---}}),
the combined indicator $c_{ab}$ (\protect\rule[0.5ex]{1.2em}{1pt})
the flow $|F_{ab}|$ (\textcolor{red}{$- \cdot -$}), 
the load $L_{ab}$ (\textcolor{magenta}{$- - -$}), 
the edge betweenness centrality $\epsilon_{ab}$  (\textcolor{brown}{\textbf{---}})
and the topological edge-connectivity $\tau_{ab}$ ($\circ$). The curves are based on the simulation of 400 
random realization of the British power grid with heterogeneous power generation.
}
\end{figure}

Comparing the ROC curves in Figs.~\ref{fig:roc1} and \ref{roc-lin-bayes-comp} shows 
that the Bayesian approach yields qualitatively similar results as the linear classifier.
Additionally, curves generated through the Bayesian approach are smoother since they 
are proper ROC curves \cite{Egan75}, i.e., their slope is monotonously decreasing.

\subsection{Critical link identification for alternative models}
\label{sec:additional-res}

\subsubsection{Critical links in the third-order model}

\begin{figure}[tb]
   \vspace{-0.5cm}
  \centerline{
    \includegraphics[height=3.8cm]{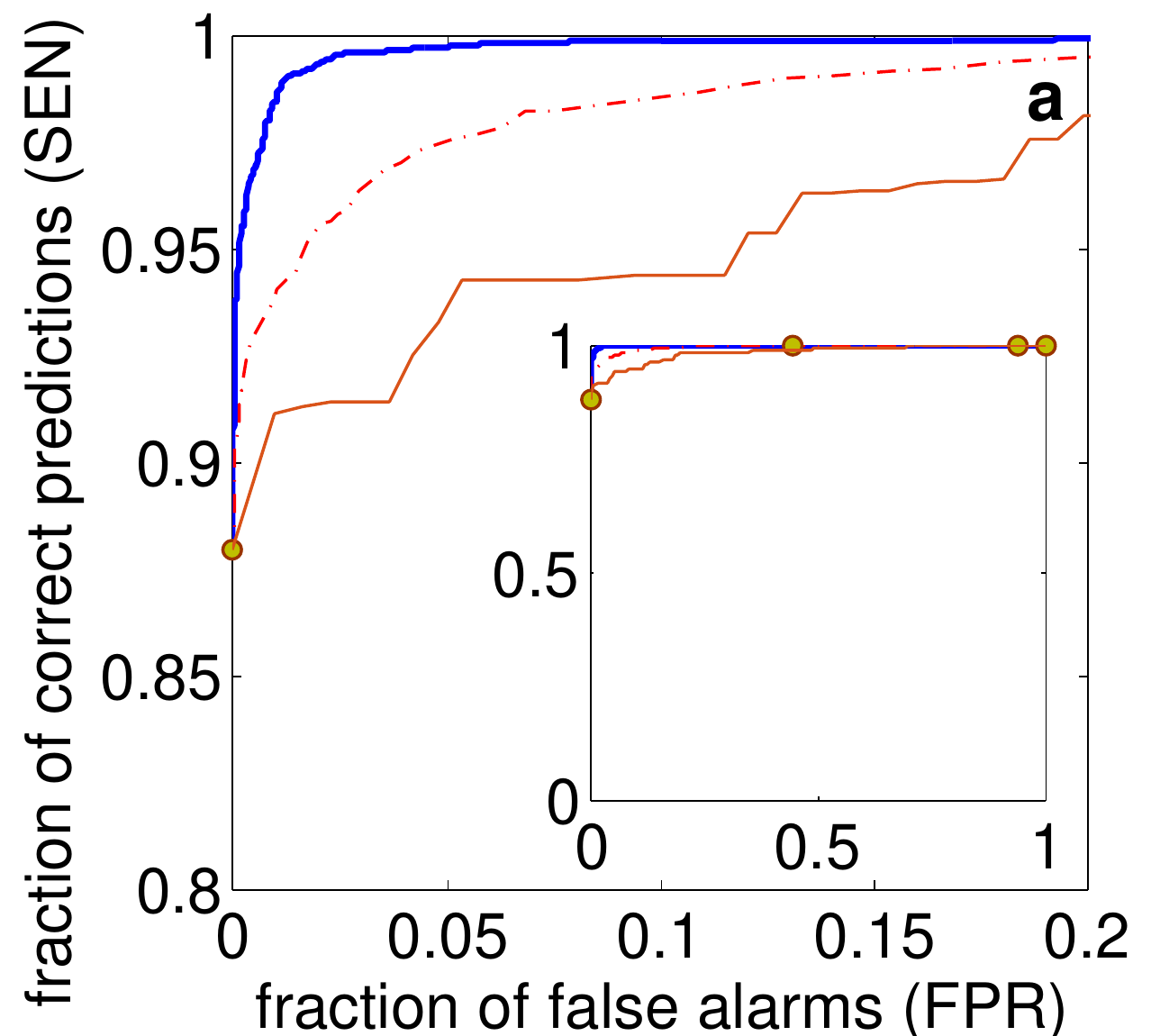}
	\includegraphics[height=3.8cm]{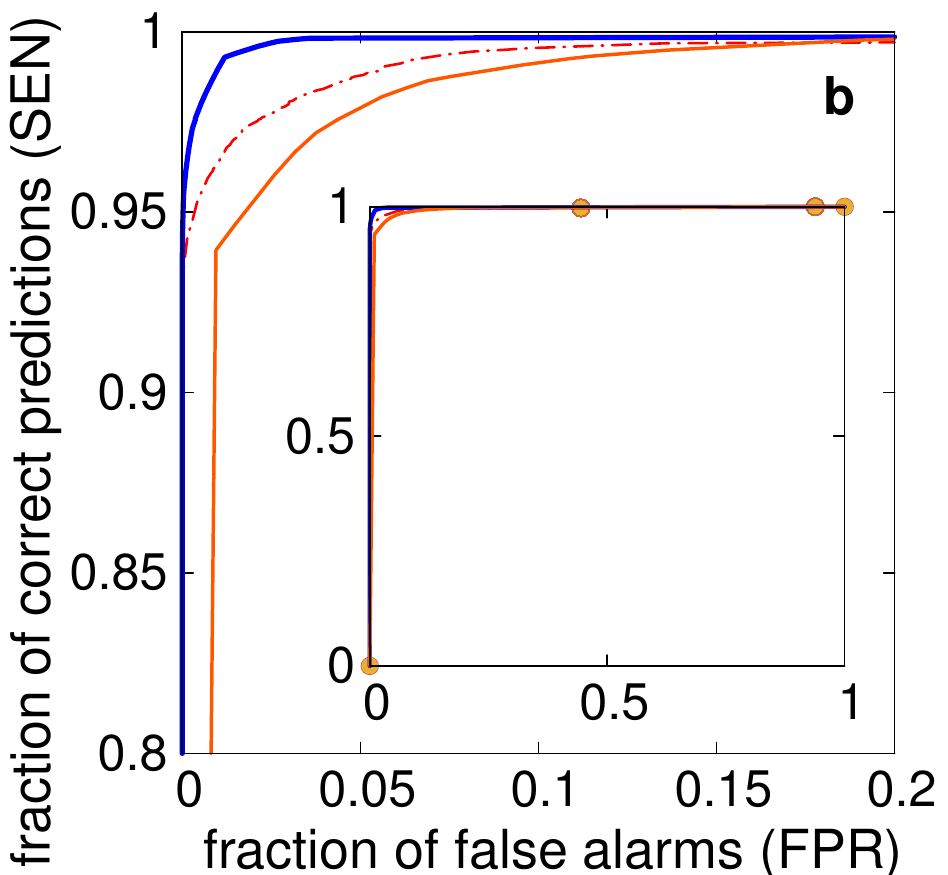}
}
\caption{
\label{fig:roc-3rdorder}
{Performance of indicators for critical links in the third order model.} 
Simulations are carried out for model power grids based on the topology of the 
British high-voltage grid (cf. Figure \ref{gbhommap}) with half of the 
nodes being generators and consumers with $P_j = \pm 1 \, {\rm s}^{-2}$. 
Shown is the ROC curve for the different indicators:
the ratio $r_{ab}$ (\textcolor{blue}{\protect\rule[0.5ex]{1.2em}{1pt}}),
the flow $|F_{ab}|$ (\textcolor{red}{$- \cdot -$}) ,
the edge betweenness centrality $\epsilon_{ab}$  (\textcolor{brown}{\textbf{---}})
and the topological edge-connectivity $\tau_{ab}$ ($\circ$). 
Results are collected for 200 random realizations of the network.
(b) ROC curves using the Bayesian prediction scheme.
}
\end{figure}

Classifying critical links based on the redundancy of the network works extremely well also for 
more complex power grid model such as the third order model introduced in section \ref{sec:3rd-order}.
We simulate the dynamics after the breakdown of a single link for random model networks
based on the topology of the British high-voltage grid (cf. Figure \ref{gbhommap}).
Half of the nodes are randomly chosen to be generators or consumers with 
$P_j = \pm 1 \, {\rm s}^{-2}$. All nodes have the nominal voltage $U_k^{(f)} = 1.1$
and we assume $B_{k \ell} = 20$ for all links and $\Delta X_k^{(d)}=1$ for all nodes
(cf.~\cite{Schm13} for the choice of parameters). 
The ROC curve obtained from the simulation of 200 random networks 
(see figure \ref{fig:roc-3rdorder}) shows that the indicator based on the redundant 
capacity significantly outperforms the indicators 
based on the flow, topological edge-connectivity
or edge betweenness centrality.

\subsubsection{Critical links in the load flow calculations}
\label{sec:res-loadflow}

Load flow calculations are routinely carried out by the grid operators 
to analyze the current state of the grid as well as the security after the 
breakdown of a single transmission line (the so-called $n-1$ case). 
We test the novel classification schemes for telling apart critical from stable links for load flow calculations, including ohmic losses and reactive power flow.

Here, we do load flow calculation for synthetic model networks based on the 
topology of the British power grid. As before we consider two scenarios for 
generation and consumption. In the homogeneous scenario, we assume that 
half of the nodes generate the power
$P_{\rm gen} = P_0$ and half are consumers with power demand 
$P_{\rm con} = -P_0$ and $Q_{\rm con} = -0.33 P_0$.
All transmission lines are assumed to have the same impedance $Z_0=0.1+i\,0.5\,\text{p.u.}$.
For the heterogeneous network, 10 of the 120 nodes are randomly chosen 
as generators ($P_{\rm gen}=11P_0$) and the rest are set to be consumers 
($P_{\rm con} = -P_0$, $Q_{\rm con} = -0.33 P_0$ as above).
The transmission lines connecting the generators to the grid are assumed to be stronger, 
i.e. to have half the impedance as the remaining ones ($Z' = 0.5 Z_0$).
For both networks the terminal voltage of the generators $|U_\text{gen}|$ 
is set to be $1\,\text{p.u.}$ and we set $P_0 = 1 \text{MW}$.
For these networks we perform load flow calculations to determine the normal 
operational state of the grid, commonly called the $n-0$ basis case. 
If no solution can be found, we discard the random realization.

We then simulate the failure of a single transmission line (the $n-1$ case)
by removing one link $(a,b)$ from the network. The post-fault condition 
of the network is then determined by applying load flow calculations again 
for the modified topology using the the pre-fault stable state as initial condition 
for the solver. The grid is considered to be stable if a solution can be found for 
the post-fault topology, and the apparent power flow $|S_{ij}|$ carried by each 
line does not exceed the security limit $K_{ij}$ of this line.
For better comparability with the previous sections we refer to $K_{ij}$ as 
the capacity of the line $(i,j)$. In other words, the grid remains stable if 
none of the other lines is overloaded when the transmission line $(a,b)$ fails.
If not, the grid is assumed to be unstable and the transmission line $(a,b)$ 
is said to be critical.

\begin{figure}[tb]
\vspace{-0.5cm}
  \centerline{
\includegraphics[width=4.4cm]{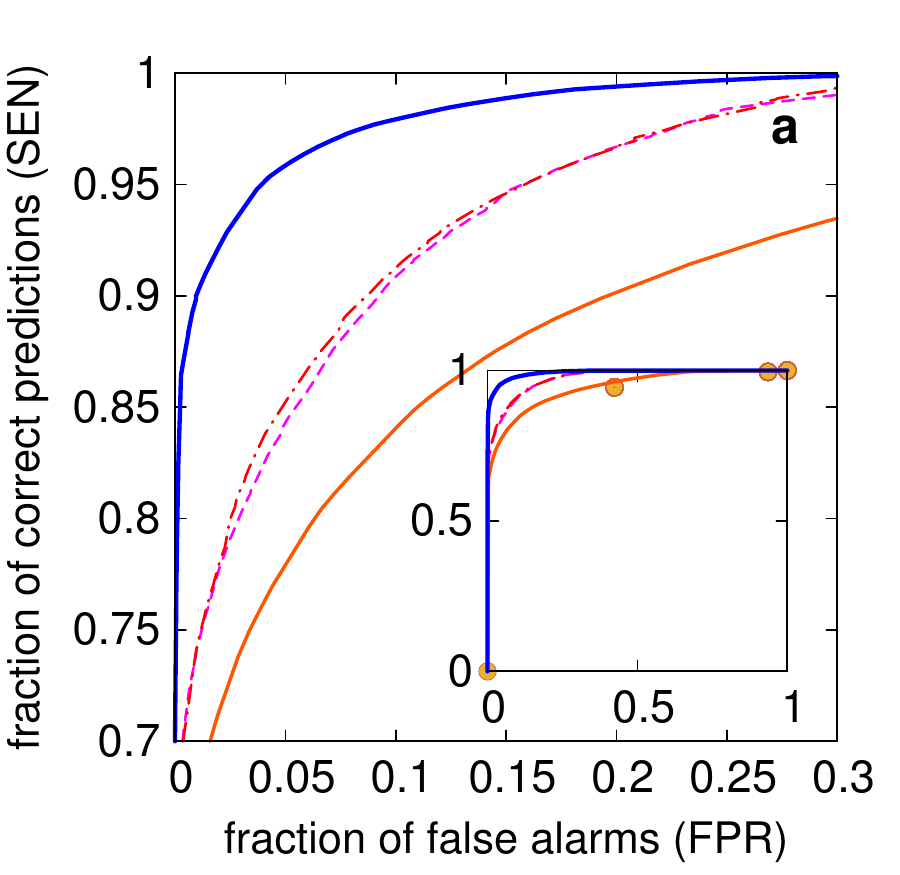}
\includegraphics[width=4.4cm]{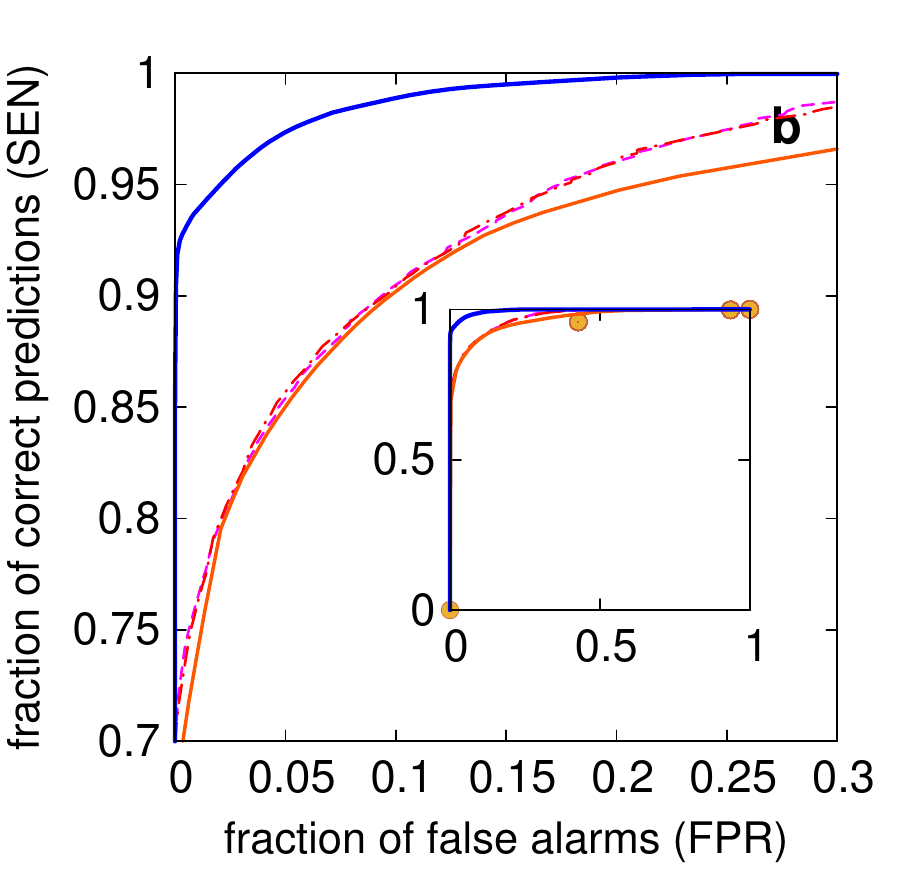}
}
\caption{\label{fig:pfroc}
{Performance of indicators for critical links in the load flow calculations.} 
Simulations are carried out for model power grids based on the topology of the 
British high-voltage grid (cf. Figure \ref{gbhommap}) for (a) the heterogeneous 
case (ten nodes randomly selected as generators) and (b) the homogeneous case 
(half of the nodes randomly chosen as generators).
Shown is the ROC curve for the different indicators: the ratio
$r_{ab}$ (\textcolor{blue}{\protect\rule[0.5ex]{1.2em}{1pt}}), 
the apparent power flow $|S_{ab}|$ (\textcolor{red}{$- \cdot -$}), the norm of 
the complex current $|I_{ab}|$ (\textcolor{magenta}{$- - -$}), and the 
topological edge-connectivity $\tau_{ab}$ ($\circ$).
ROC curves are calculated using the Bayesian prediction scheme described in 
Sec.~\ref{sec:adv_stat} using data from 100 random realizations of the network.
Bridges are considered trivially critical.}
\end{figure}

In load flow calculations the definition of the redundant capacity of a transmission 
line is slightly different from the one for the oscillator model, since the power flow 
between nodes is complex. For the homogeneous network, the transmission capacity 
is identical for all links, which is defined as $1.2$ times the maximal apparent power 
flow in the stable operation state in the $n-0$ case, 
\be
   K^{\text{LF}} :=1.2 \times  \max_{m,n} |S^{(n-0)}_{mn}|.
\ee
For the heterogeneous network, the transmission capacity of the links connected 
to the generators $K_\text{gen}^{\text{LF}}$ is set as $2$ times the capacity of the links 
connecting only consumers $K_\text{con}^{\text{LF}}$: $K_\text{gen}^{\text{LF}}=2K_\text{con}^{\text{LF}}$, 
where $K_\text{con}$ is defined regarding the maximal apparent power flow 
on the consumer connecting lines maximal apparent power flow in the stable 
operation state in the $n-0$ case,
\be 
   K_{\text{con}}^{\text{LF}} :=1.2 \times  \max_{\text{consumers} \,  m,n} |S^{(n-0)}_{mn}|.
\ee
By this definition, each transmission line in the stable operation state of the 
initial pre-fault networks is working safely without any overload. The redundant 
capacity of a transmission line $K^\text{red}_{ab}$ is then defined by the 
algorithm introduced in section \ref{sec:kred}.
The only difference is that the flow $F_{ij}$ is replaced by the apparent power flow 
$|S_{ij}|$ times the sign of the active power flow $P_{ij}$, which indicates the 
direction of the net power flow between nodes, i.e.
\be
	F_{ij}^{\text{LF}} :=\text{sgn}(P_{ij})\cdot|S_{ij}|.
\ee

\begin{figure}[t!]
\centering
\includegraphics[width=4.2cm]{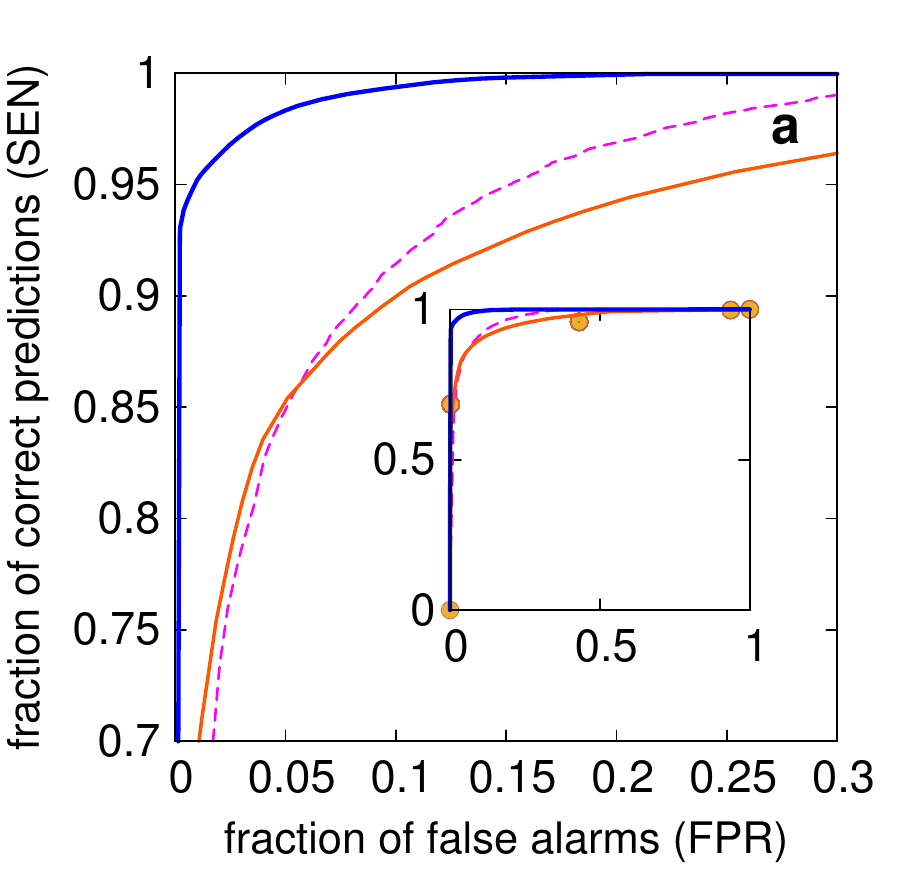}
\includegraphics[width=4.2cm]{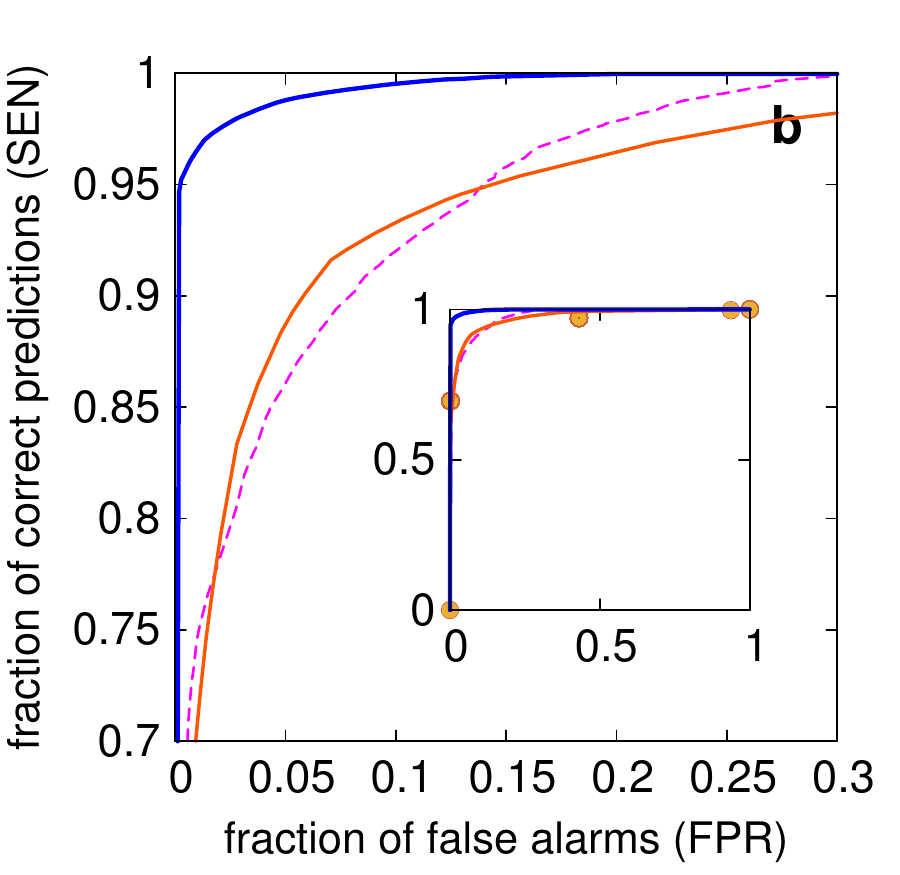}
\caption{\label{fig:dcroc}
{Performance of indicators for critical links in the linear flow model.} Simulations 
are carried out for model power grids based on the topology of the British high-voltage grid 
(cf. Figure \ref{gbhommap}) for (a) the heterogeneous case (ten nodes randomly selected 
as generators) and (b) the homogeneous case (half of the nodes randomly chosen as 
generators). Shown is the ROC curve for the different indicators: the ratio $r_{ab}$ 
(\textcolor{blue}{\protect\rule[0.5ex]{1.2em}{1pt}}), the initial flow $|F_{ab}|$ 
(\textcolor{magenta}{$- - -$}), and the topological  edge-connectivity $\tau_{ab}$ 
($\circ$). ROC curves are calculated using the Bayesian prediction scheme described in 
Sec.~\ref{sec:adv_stat} using data from 100 random realizations of the network.
}
\end{figure}

As indicators for the criticality of a transmission line $(a,b)$, we test the apparent 
power flow $|S_{ab}|$, the norm of the complex current $|I_{ab}|$, the ratio 
$r_{ab}=|F_{ab}^{\text{LF}}/K_{ab}^{\text{red}}|$, 
the topological edge-connectivity $\tau_{ab}$
and the edge betweenness centrality $\epsilon_{ab}$.
The novel indicator based on the 
redundant capacity $r_{ab}$  shows a much better predictive 
performance than other indicators, see Figure \ref{fig:pfroc}. 

\begin{figure*}[tb]
  \centerline{
    \includegraphics[width=5.5cm, angle=0]{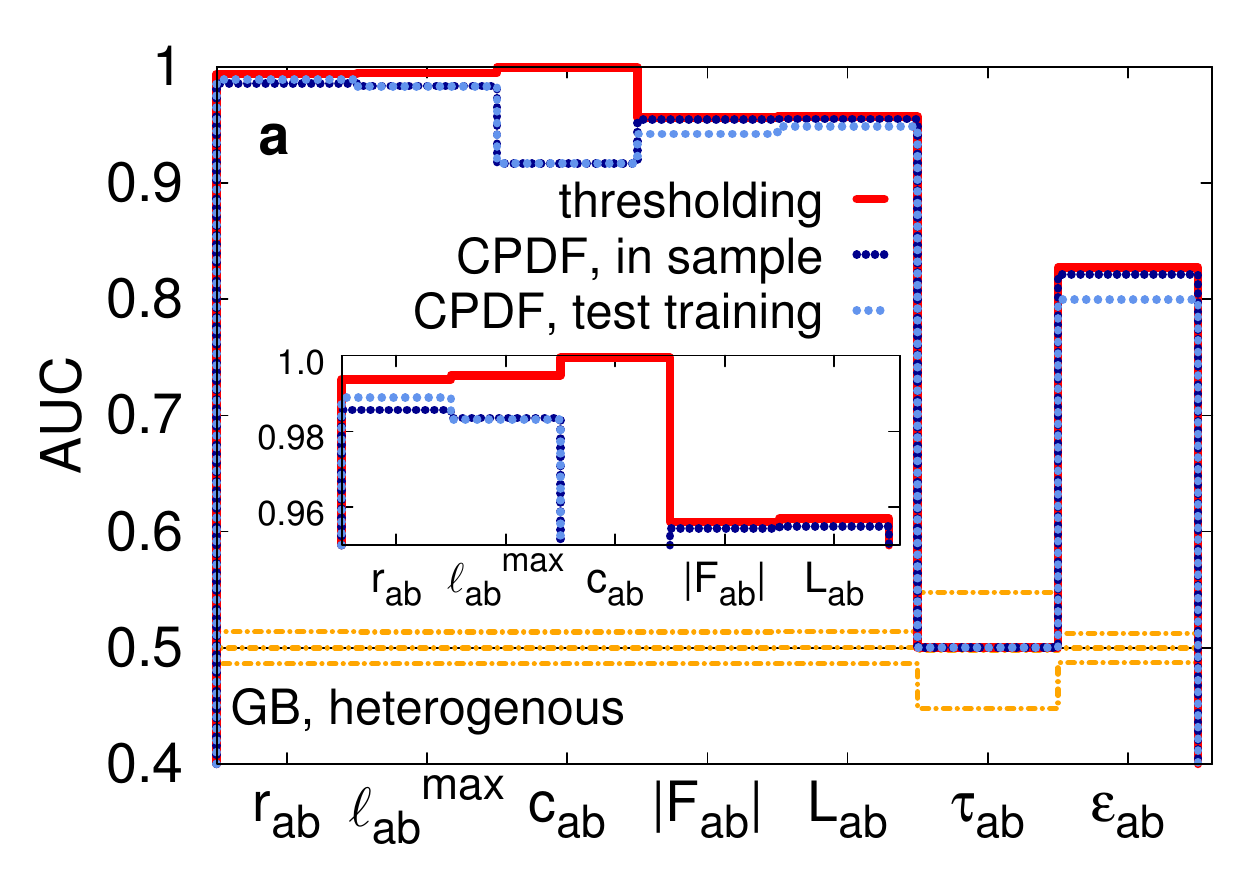}
    \includegraphics[width=5.5cm, angle=0]{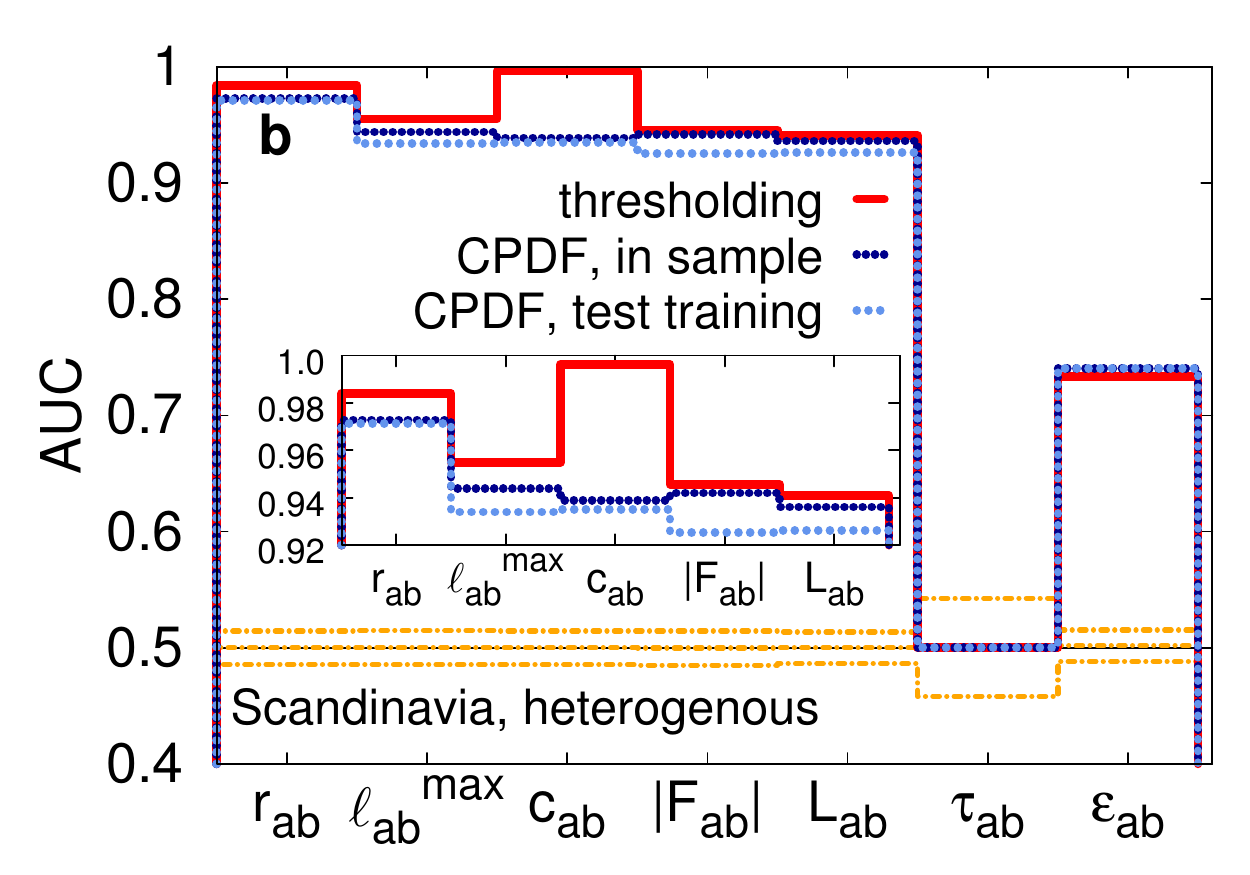}
    \includegraphics[width=5.5cm, angle=0]{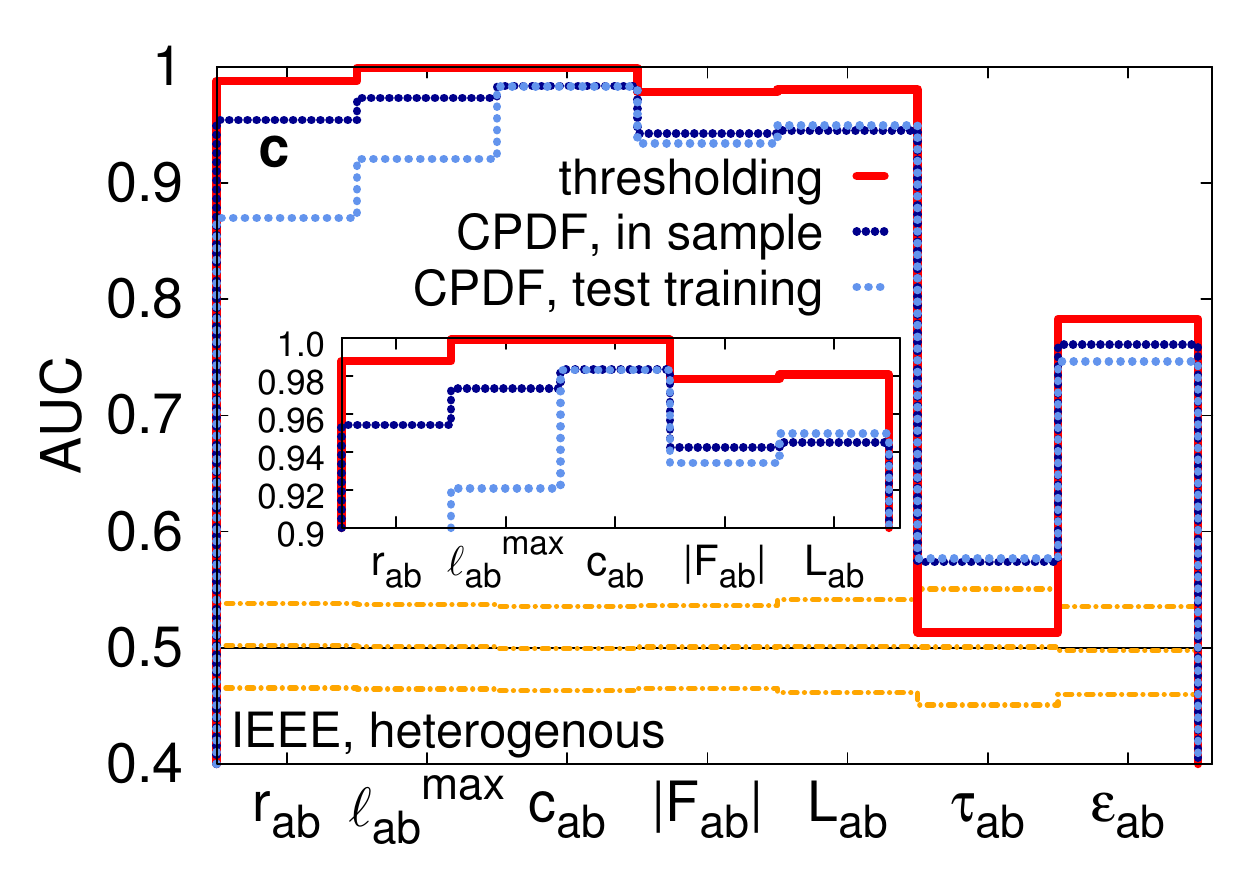}
  }
  \centerline{
    \includegraphics[width=5.5cm, angle=0]{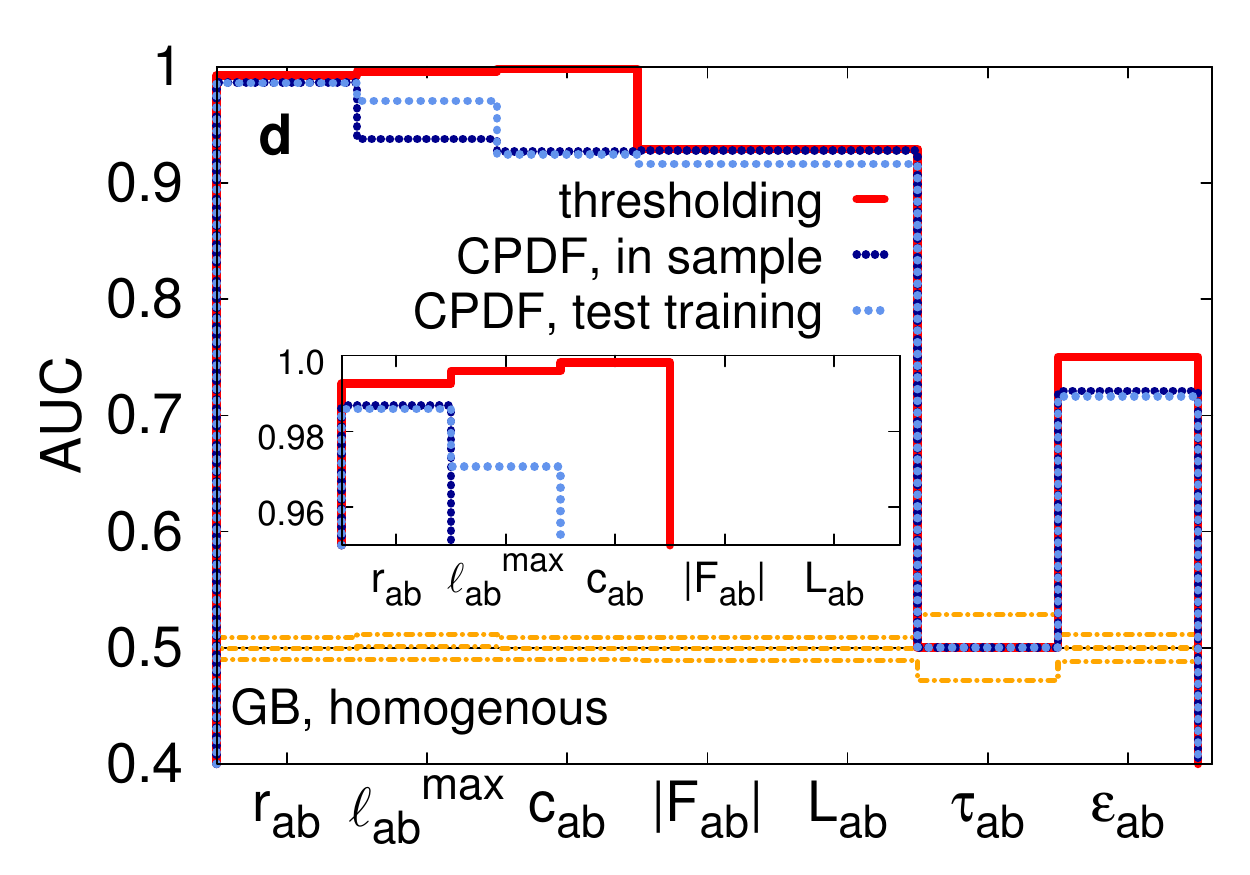}
    \includegraphics[width=5.5cm, angle=0]{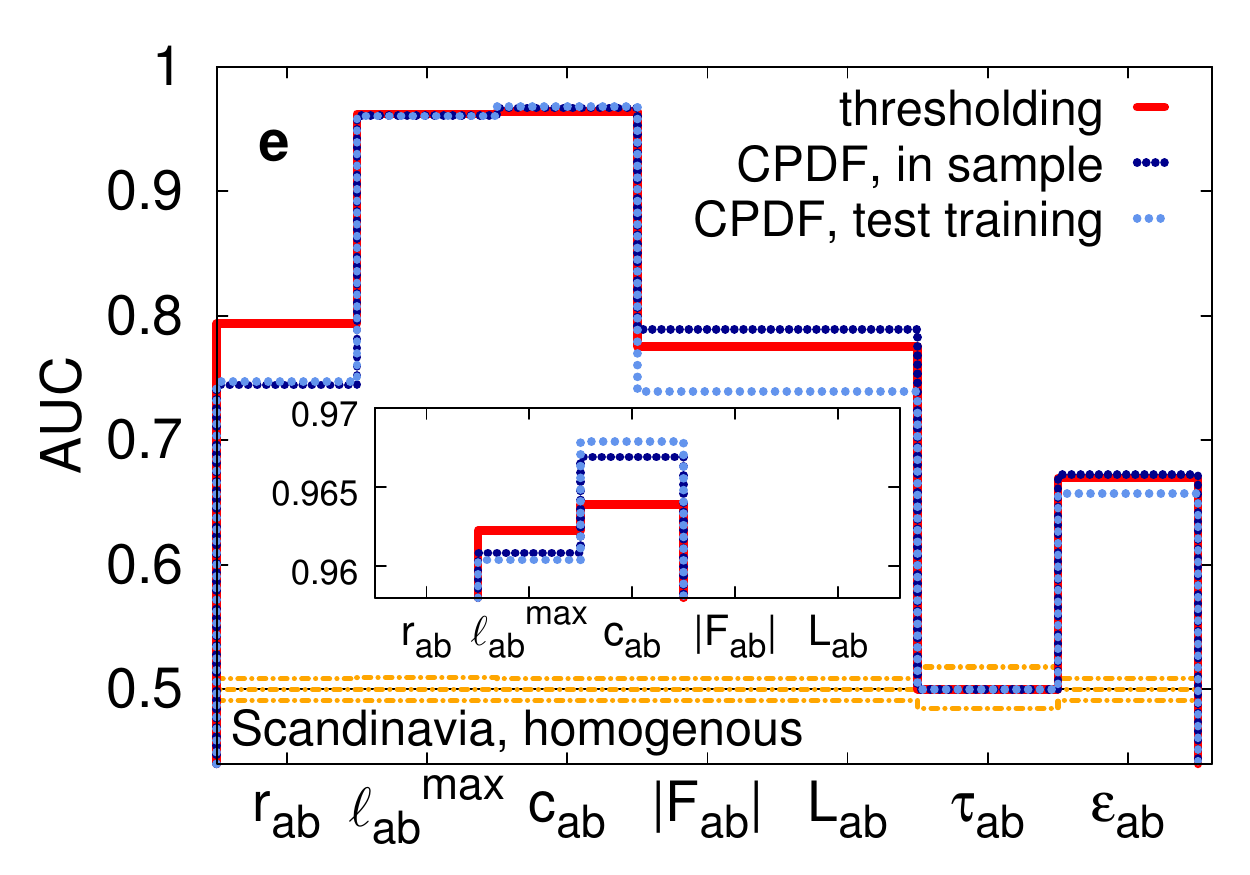}
    \includegraphics[width=5.5cm, angle=0]{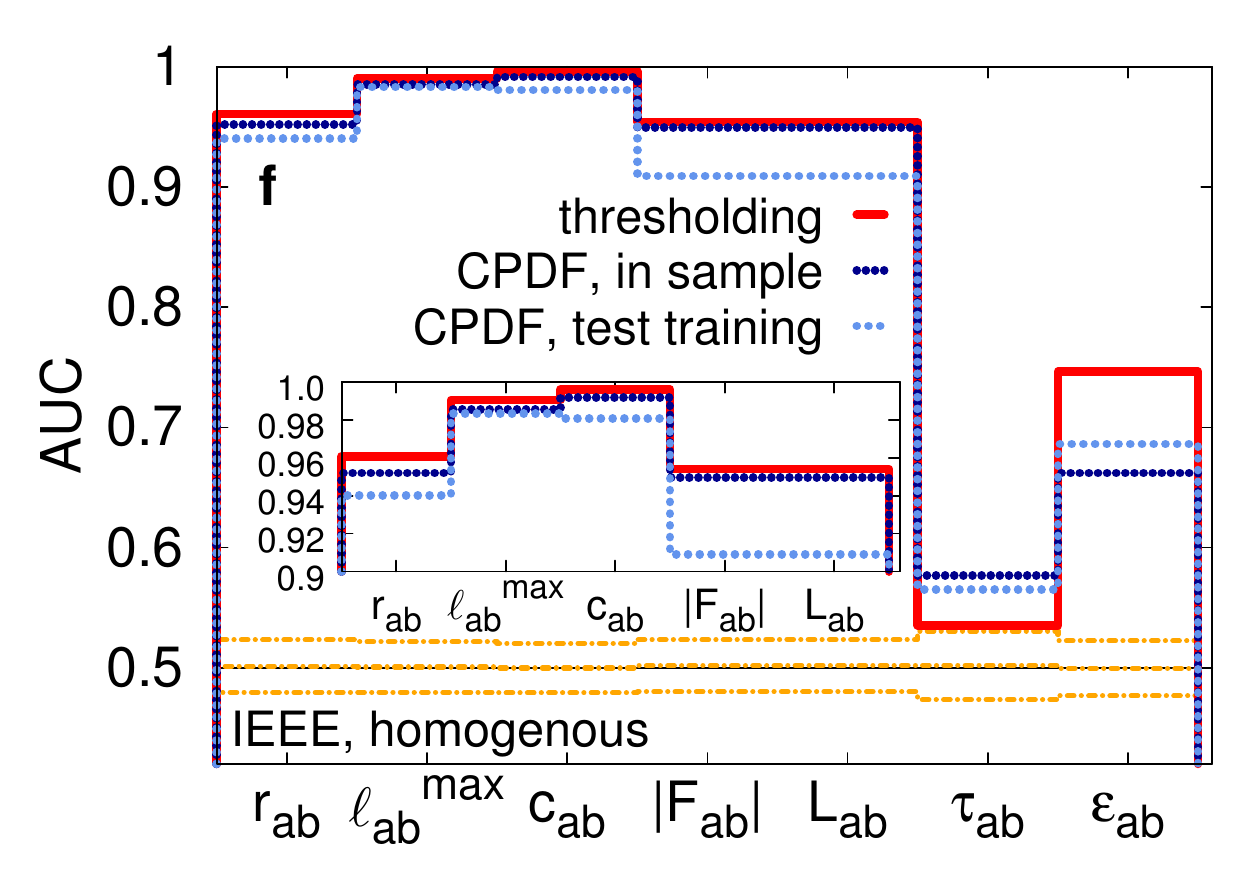}
  }
\caption{\label{auc-lin-bayes-comp}
{Summary of the performance of different indicators for various network topologies.}
Shown is the AUC for critical links using the linear classification scheme (red solid lines), 
the Bayesian prediction scheme in sample (dark blue, dashed) and the Bayesian prediction 
scheme out of sample with a separation of test and training data set (light blue dashed lines). 
Confidence bounds (yellow) were estimated making random predictions.
Data averaged over 400 random realizations for the British and the IEEE grid and 
200 for the Scandinavian grid, respectively.
}
\end{figure*}

\subsubsection{Critical links in the linear flow model}

The linear flow model can be considered as the DC approximation of the load 
flow calculations. Therefore, the simulation procedures and the parameter 
settings are basically the same, except for the following differences: 
(1) The magnitude of the voltage is fixed for all nodes, while the reactive 
power $Q_k$ is arbitrary. 
(2) Ohmic losses are neglected such that the admittance of the transmission 
lines is purely imaginary, $Y_0=iB=i2\,\text{p.u.}$.
(3) The flows are defined as
\be
   F^{\text{LiF}}_{ij}:=B_{ij}(\theta_i-\theta_j).
\label{eq:lifeq}
\ee

Similar as in the load flow calculations, a transmission line is considered as 
critical when its failure causes a secondary overload of at least one other 
transmission lines. The overload of a line is defined as following: the maximal 
flow in the network exceeds the line capacity, which is defined for the linear 
flow model as
\be
K^{\text{LiF}}:=1.2 \times  \max_{m,n} |F^{(n-0)}_{mn}|
\ee
for the homogeneous network and 
\begin{align}
   K_{\text{con}}^{\text{LiF}}&:=1.2 \times  \max_{\text{consumers} \,  m,n} |F^{(n-0)}_{mn}|,\\
   K^{\text{LiF}}_\text{gen}&:=2K^{\text{LiF}}_\text{con}
\end{align}
for the heterogeneous networks.

The indicators we test include the magnitude of the initial flow of the lines 
$|F_{ab}|$, the ratio of the initial flow and the redundant capacity 
$r_{ab}=|F_{ab}/K^{\text{red}}_{ab}|$, the topological edge-connectivity 
$\tau_{ab}$ and the edge betweenness centrality $\epsilon_{ab}$.
The indicator $r_{ab}$ based on redundant capacity significantly 
outperforms the other indicators (Figure \ref{fig:dcroc}).

\subsection{Critical link identification on alternative topologies}
\label{sec:add-topologies-results}

We test the performance of the novel indicators by extensive numerical 
simulations of the power grid dynamics for various network topologies, 
as described in Sec.~\ref{sec:networkdata}, using the oscillator model.
In Fig.~\ref{auc-lin-bayes-comp} we compare AUCs for different indicators 
and different ways of prediction: making random predictions, using a 
indicator as a linear classifier, using a Bayesian classifier for in-sample 
prediction and using a Bayesian classifier with separation of test and 
training data set. In the later case, CPDFs were evaluated using $50 \%$ 
of the available data and ROC curves were generated using the remaining $50\%$.
The composition of test and training data set in terms of numbers of 
bridges, stable and critical links was chosen randomly.
Mostly, linear and Bayesian approach yield comparable results, with the 
linear classifier being slightly better. This is due to the fact that the relation 
between the value of the indicator variable and its prediction success is in many 
cases simple, such that it can be capture by a linear classifier.
Estimating this relation based on CPDFs on the other hand can be difficult, 
if the data is clustered around one value or spread around several orders of magnitude.
Comparing prediction success, the novel indicators $r_{ab} = |F_{ab}/K_{ab}^{\rm red}|$, 
$\ell^{\rm max}$ and $c_{ab}$ perform better than common indicators  in all cases studied.

A closer inspection reveals that the predicted maximum load $\ell^{\rm max}$ 
performs better than the ratio $r_{ab}$ for the datasets with homogeneous 
supply, i.e. equal number of generators and consumers.
One reason for this finding could be the fact that the total number of 
critical links is higher for homogeneous than for heterogeneous supply.
Furthermore we find that the performance of all indicators is lowest 
for the Scandinavian grid with homogeneous supply.
This model network is the one with the highest fraction of critical links in total.

% --- Literatur -------------------------------------------------------------------

%\bibliography{redundancy}
%\bibliographystyle{apsrev}

\end{document}